\documentclass[journal]{IEEEtran}
\pdfoutput=1
\ifCLASSINFOpdf
\else
\fi

\usepackage{cite}
\usepackage{booktabs}
\usepackage{multirow}
\usepackage{amsmath,amssymb,amsfonts,amsthm}
\usepackage[ruled, linesnumbered]{algorithm2e}
\usepackage{textcomp}
\usepackage{url}
\usepackage{bbm}
\usepackage{enumerate}
\usepackage{xcolor}
\usepackage{graphicx}
\usepackage{graphics}
\usepackage{subfigure}
\usepackage{pifont}

\newtheorem{theorem}{Theorem}
\ifodd 1
\newcommand{\com}[1]{\textbf{\color{red} (COMMENT: #1)}} %comment of the text
\newcommand{\comg}[1]{\textbf{\color{green} (COMMENT: #1)}}
\newcommand{\response}[1]{\textbf{\color{magenta} (RESPONSE: #1)}} %response to comment
\else

\newcommand{\com}[1]{}
\newcommand{\comg}[1]{}
\newcommand{\response}[1]{}
\fi
\begin{document}

\title{Serving Graph Neural Networks With Distributed Fog Servers For Smart IoT Services}

\author{Liekang~Zeng,
        Xu~Chen,
        Peng~Huang,
        Ke~Luo,
        Xiaoxi~Zhang,
        and~Zhi~Zhou
\thanks{
The authors are with the School of Computer Science and Engineering, Sun Yat-sen University, Guangzhou, Guangdong, 510006 China (e-mail: zenglk3@mail2.sysu.edu.cn, chenxu35@mail.sysu.edu.cn, \{huangp57, luok7\}@mail2.sysu.edu.cn, \{zhangxx89, zhouzhi9\}@mail.sysu.edu.cn).
}
}

\maketitle

\begin{abstract}
    Graph Neural Networks (GNNs) have gained growing interest in miscellaneous applications owing to their outstanding ability in extracting latent representation on graph structures. 
    To render GNN-based service for IoT-driven smart applications, traditional model serving paradigms usually resort to the cloud by fully uploading geo-distributed input data to remote datacenters.
    However, our empirical measurements reveal the significant communication overhead of such cloud-based serving and highlight the profound potential in applying the emerging fog computing.
    To maximize the architectural benefits brought by fog computing, in this paper, we present Fograph, a novel distributed real-time GNN inference framework that leverages diverse and dynamic resources of multiple fog nodes in proximity to IoT data sources. 
    By introducing heterogeneity-aware execution planning and GNN-specific compression techniques, Fograph tailors its design to well accommodate the unique characteristics of GNN serving in fog environments.
    Prototype-based evaluation and case study demonstrate that Fograph significantly outperforms the state-of-the-art cloud serving and fog deployment by up to 5.39$\times$ execution speedup and 6.84$\times$ throughput improvement.
\end{abstract}


\IEEEpeerreviewmaketitle

\section{Introduction}\label{sec:introduction}

\IEEEPARstart{G}{raphs} are ubiquitous. 
Given the intuitionistic abstraction on relational structures, graphs drive the organization and computation of miscellaneous real-world data such as traffic sensory networks \cite{ye2020build, HuangHLDK20}, online social graphs \cite{fan2019graph, guo2020deep}, and power grids \cite{deka2020graphical, owerko2018predicting}.
To facilitate deep learning using such form of data, recent advances in neural networks have extrapolated to the graph domain, resulting in a new stream of models called Graph Neural Networks (GNNs).

GNNs differ from traditional Deep Neural Networks (DNNs) by integrating graph embedding techniques with convolutions \cite{zhou2020graph, abadal2021computing, besta2022parallel}.
In essence, GNNs leverage an iterative aggregation to an input graph and, through neural network operators, to capture hierarchical patterns from subgraphs of variable sizes.
This enables the model to learn the properties for specific vertices, edges, or the graph as a whole, and generalize to unobserved graphs.
Benefited from such powerful expressiveness, GNNs achieve superior prediction performance in various graph-related tasks, and have emerged as a powerful data-driven tool for enabling a multitude of real-world IoT-driven smart applications, \textit{e.g.}, traffic flow forecasting \cite{li2017diffusion, wang2020traffic}, location-based recommendation \cite{zhong2020hybrid, chang2020learning}, and vehicle trajectory prediction \cite{jeon2020scale, zhou2021ast}.

\begin{figure}[t]
  \centering
  \includegraphics[width=0.9\linewidth]{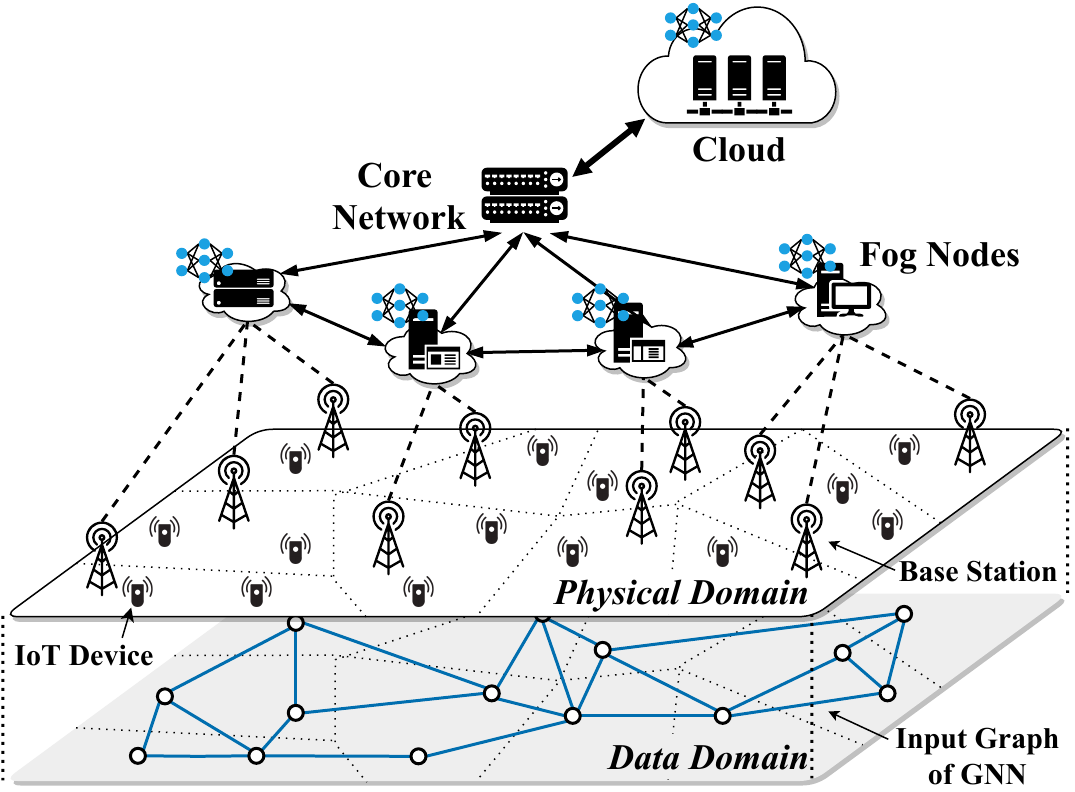}
  \caption{An example scenario of GNN serving at the network edge.
  Instead of entirely offloading IoT sensory data to the cloud via the delay-significant Internet, Fograph processes GNN workloads over fog nodes in proximity to enable real-time GNN serving.}
  \label{fig:scenario}
\end{figure}

To render smooth services for these applications, the \textit{de facto} standard methodology is to offload raw data and computation to central cloud servers \cite{zhu2019aligraph, zhang2020agl}.
For instance, in Fig. \ref{fig:scenario}, the massive sensory data from IoT devices are fully uploaded (in physical domain) and their corresponding GNN input graph (in data domain) is computed at a remote cloud.
While this paradigm may act well for many CNN-based image processing tasks \cite{crankshaw2020inferline, crankshaw2017clipper, gujarati2020serving}, however, it can reach suboptimal performance for GNN model serving due to its unique input characteristics.
First, the input graph of GNN typically spans geographically with scattered data sources, \textit{e.g.} the IoT sensory devices, as vertices. 
Unlike image or video from a single source, to obtain the complete input for one inference, GNN execution is obliged to wait until all correlated data points arrive, which considerably prolongs the total serving latency.
Second, as the graph scales and the number of vertices increases, the input data size grows linearly and can become much larger than an ordinary CNN inference input, emphasizing the communication stress.
Worse still, the transmission cost is further magnified due to the long transmission delay of Wide Area Network (WAN) and potential network congestion.
Specifically, as we will show later in \S\ref{sec:exam_serve}, the data uploading phase could dominate the whole procedure by consuming $>$95\% latency in a typical cloud-based GNN serving.

To tame such intractability, a promising solution is to exploit available computing resources in proximity to data sources with the emerging fog computing\footnote{In the terminology of some literature, \textit{fog computing} also refers to \textit{edge computing}. Since \textit{edge} has represented the links in graphs, throughout this paper, we use the term \textit{fog} to avoid ambiguity.} paradigm \cite{zhou2019edge}.
Concretely, as Fig. \ref{fig:scenario} illustrates, we can sink GNN workload from remote cloud into vicinal fog nodes\footnote{We exclusively use \textit{node} to denote a fog server and leave \textit{vertex} for graphs.} (\textit{e.g.} 5G fog servers) and manage data collection and computation within the Local Area Network (LAN).
Consequently, the avoidance of unreliable WAN connections allows observably lower communication overhead, reducing at most 67\% data collection latency in our experimental measurements. 
In brief, fog computing exhibits prospective potential for real-time GNN serving at the network edge.

Nevertheless, despite the advantages, efficient fog deployment yet suffers from a set of challenges.
First, different from the cloud that takes computing resources as a whole, the fog environments usually consist of loosely coupled nodes \cite{wang2020convergence}.
To adapt complex GNN processing over them, a distributed counterpart is required, where input data need to be judiciously placed and routed to respective fog nodes for distributed execution.
Second, fog environments are inherently heterogeneous \cite{shi2016edge}, \textit{e.g.} with computing facilities ranging from small-size gateways \cite{chen2018edge} to powerful cloudlets \cite{satyanarayanan2009case}; their available bandwidth allocated for serving also vary.
To exploit the maximum parallelization from the diversity, a heterogeneity-aware data placement strategy with effective load balancing is highly desired.
Further complicating the problem is dynamic factors like fog nodes' load levels, network conditions, \textit{etc.}, which may dramatically decline the performance of the whole pipeline.
Unfortunately, existing GNN serving mechanisms cannot sufficiently meet these requirements.

To this end, we present Fograph, a novel distributed system that enables real-time GNN inference over multiple heterogeneous fog nodes.
Fograph's contribution goes beyond merely applying fog computing to boost GNN serving, instead it addresses the above challenges at four levels.
First, from an execution perspective, a holistic distributed workflow is introduced for enabling fog nodes to collaboratively serve GNN inference.
Second, to attain efficient runtime, an inference execution planner is designed to optimize the data placement of the input graph, along with a GNN-oriented profiling methodology that allows accurately characterizing heterogeneous computing capabilities.
Third, to alleviate the communication bottleneck and ameliorate the overall performance, a novel graph data packing technique is applied that leverages the topological properties and compresses transferred data with minimal impact on accuracy. 
Finally, to adapt to dynamic changes such as load fluctuation, a dual-mode workload scheduler is developed which progressively adjusts the graph data placement in order to acquire the best-performing configuration.
Extensive evaluation against multiple benchmarks demonstrates Fograph's superior performance gain over traditional cloud serving and straw-man fog counterpart.
In summary, this work makes the following key contributions.
\begin{itemize}
    \item An empirical study on GNN serving latency with existing cloud and basic fog mechanisms. By breaking down the overheads of communication and execution, we observe a major cost reduction on the communication side, highlighting utilizing fog computing as a promising optimization opportunity (\S\ref{sec:background}).
    \item A regularized workflow for fog-enabled GNN serving that covers the full lifecycle from offline configuration to online data collection and distributed runtime. Data parallelism is applied and retrofitted by leveraging the execution characteristics of GNN inference in order to collaborate multiple fog nodes (\S\ref{sec:execution}).
    \item A heterogeneity-aware distributed GNN inference system Fograph that enables real-time performance. Reflecting on the diverse and fluctuated fog resources, we design an inference execution planner with load balancing for maximum parallelization (\S\ref{sec:metadata}, \S\ref{sec:planning}), and a dual-mode workload scheduler to accommodate dynamics (\S\ref{sec:adapter}). 
    \item A GNN-specific packing mechanism that exploits the reduced-precision resilience and sparsity of GNN workloads to minimize data uploading overhead. Our communication optimizer combines a lossless compressor and a degree-aware lossy quantizer, which exposes previously unattainable designs for distributed GNN inference, while not sacrificing the predition accuracy of the system (\S\ref{sec:compression}).
    \item A comprehensive evaluation of Fograph using multiple benchmarks, demonstrating its superiority over state-of-the-art cloud serving and straw-man fog deployment by up to 5.39$\times$ execution speedup and 6.84$\times$ throughput improvement (\S\ref{sec:evaluation}).
\end{itemize}

\section{Background and Motivation} \label{sec:background}

\subsection{Graph Neural Networks}
\label{sec:gnn}

\begin{figure}[t]
    \centering
    \subfigure[Input graph.]{
        \begin{minipage}[t]{0.24\linewidth}
        \centering
        \includegraphics[height=2.1cm]{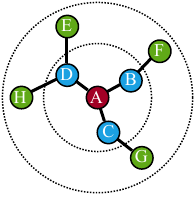}
        \label{fig:gnn_input_graph}
        \end{minipage}
    }
    \subfigure[Inference process.]{
        \begin{minipage}[t]{0.68\linewidth}
        \centering
        \includegraphics[height=2.4cm]{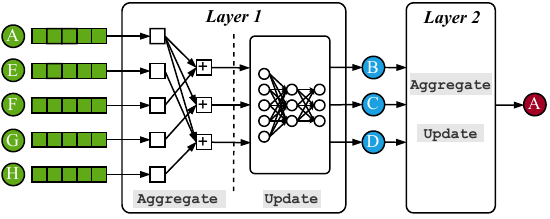}
        \label{fig:gnn_inference}
        \end{minipage}
    }
    \caption{An GNN inference instance of computing vertex \textit{A}'s embedding through two GNN layers. }
    \label{fig:gnn}
\end{figure}

The real-world graphs typically contain two kinds of data.
One is the adjacency matrix, implying the global structural information, and the other is the feature vectors that describe vertices and edges' physical properties.
GNNs take both data as input and learn a representation vector, called \textit{embedding} or \textit{activation}, for each vertex.
The learned representation can be used for downstream tasks such as vertex clustering, link prediction, and graph classification \cite{zhou2020graph}.

In Fig. \ref{fig:gnn}, we illustrate a two-layer GNN instance from the perspectives of data flow during inference process.
Fig. \ref{fig:gnn_input_graph} depicts the input graph, with vertex $A$ and its one-hop and two-hop neighbors in different color.
Fig. \ref{fig:gnn_inference} unfolds a layer's detail operations.
Essentially, each GNN layer collectively aggregates the neighbor vertices' activations from the previous layer's output, and then updates the target vertex's activation using a neural network operator such as convolution or multi-layer perceptron.
Within the same layer, all vertices share the same weights in \texttt{Aggregate} and \texttt{Update} functions, while different layers may differ them.
To compute embeddings through a $K$-layer GNN, vertices should retrieve information from their $K$-hop neighbors.
Formally, the computation of the $k$-th GNN layer on vertex $v$ can be described as:
\begin{align}
    a^{(k)}_v &= \texttt{Aggregate}(\{h^{(k-1)}_u|u \in \mathcal{N}_v\} ), \\
    h^{(k)}_v &= \texttt{Update}(a^{(k)}_v, h^{(k-1)}_v),
\end{align}
where $h^{(k)}_v$ is the representation vector of vertex $v$ at the $k$-th layer, $h^{(0)}_v$ is initialized by the input features of $v$, and $\mathcal{N}_v$ denotes the vertices set of $v$'s direct neighbors.

\begin{table}[t]
    \caption{Example GNN inference functions.}
    \label{table:models}
    \centering
    \begin{tabular}{|r|l|}
    \hline
    \textbf{Model} & \textbf{Functions} \\ \hline \hline
    \multirow{2}{*}{GCN}       
    & $ a^{(k)}_v = \sum\limits_{u\in\mathcal{N}_v} h^{(k-1)}_u,$  \\
    & $h^{(k)}_v = \sigma (W^{(k)} \cdot \frac{a^{(k)}_v+h^{(k-1)}_v}{|\mathcal{N}_v|+1}).$   \\ \hline
    \multirow{2}{*}{GAT}       
    & $a^{(k)}_v = \sum\limits_{u\in\mathcal{N}_v\cup \{v\}} \alpha^{(k)}_{vu} W^{(k)} h^{(k-1)}_u,$  \\
    & $h^{(k)}_v = \sigma( a^{(k)}_v ).$\\ \hline
    \multirow{2}{*}{GraphSAGE} 
    & $ a^{(k)}_v = \frac{1}{|\mathcal{N}_v|}\sum\limits_{u\in\mathcal{N}_v} h^{(k-1)}_u,$ \\
    & $h^{(k)}_v = \sigma( W^{(k)} \cdot (a^{(k)}_v, h^{(k-1)}_v) ).$ \\ 
    \hline
    \end{tabular}
\end{table}

\textbf{Examples.}
Table \ref{table:models} lists three popular GNN models to exemplify the above two functions.
GCN \cite{kipf2016semi} is one of the first graph learning models that bridge the gap between spectral transformation and spatial convolutions.
Its \texttt{Aggregate} simply uses a summation and the \texttt{Update} puts weighted aggregation to an element-wise nonlinearity $\sigma(\cdot)$.
GAT \cite{velivckovic2017graph} is the representative of another GNN category that incorporates attention mechanism into feature propagation.
Its inference directly uses the learned attention parameters $\alpha^{(k)}_{vu}$ to weight neighbors and passes the aggregation through the nonlinearity for output.
GraphSAGE \cite{hamilton2017inductive} is recognized as the classic inductive GNN variant.
While its training adopts sampling-based techniques to trade accuracy with training speed, its inference fully collects the neighbor sets for aggregation and update.
Here in Table \ref{table:models} we formalize its mean aggregate version.

\subsection{Emerging Real-Time GNN Applications}
\label{sec:application}

GNNs have been submerged in many scenarios with real-time responsiveness requirements, particularly for many emerging IoT-enabled smart applications.
In the following, we provide several motivating examples.

\textbf{Traffic flow forecasting.}
Accurately forecasting traffic speed, volume or the density of roads is the fundamental problem in Intelligent Transportation Systems (ITS).
To support such intelligence, GNNs construct spatial-temporal models to perform graph-level predictions.
For instance, some models \cite{li2017diffusion, guo2019attention, yu2017spatio} consider traffic sensory networks as a spatial-temporal graph where the vertices are roadside detectors and the edges are roads.
Each vertex is attached with a time-varying vector that records immediate properties such as traffic speed and occupancy.
In such circumstances, timely prediction is of paramount importance given the speedily changing traffic and its publicly wide impacts, which requires real-time GNN processing.

\textbf{Location-based recommendation.}
Recommending yet-unvisited Point of Interest (POI) to potentially interested users has been a core function for many commercial mobile applications (\textit{e.g.} Airbnb, TripAdvisor).
To utilize the rich semantic information like geographical constraints and social influences, a number of works \cite{zhong2020hybrid, chang2020learning, yuan2020spatio} have built upon GNN models.
Several graphs are created in these systems, including a spatial graph of geo-distributed POIs, a social graph of users, and a bipartite graph connecting POIs and users based on historical consuming records.
Such kind of services typically exhibit a soft-realtime necessity - if the recommendation comes late, the results can be out of date as the user may have moved to other locations.
In other words, low latency GNN inference is demanded for rendering effective user experience.

\textbf{Vehicles trajectory navigation.}
Autonomous robotics has become a hot spot in recent years, and efficient and collision-free trajectory navigation is a key technology to ensure its mobility \cite{li2020message}.
As an example, in precision agriculture \cite{tokekar2016sensor, radoglou2020compilation}, a fleet of autonomous drones move along fields to measure and maintain the crops' health, spraying pesticides if there is an infestation.
GNN-based methods \cite{tolstaya2020learning, li2020message} enhance this procedure by mapping the vehicles as graphs and performing inference to help plan paths instantaneously.
Each drone, as a vertex, captures sensory data (for flight height, ambient light intensity, \textit{etc.}) every few seconds as features.
Any delay of the control may result in catastrophic crashes of the vehicles, for which fast inference is needed.

\subsection{Examining GNN Serving Pipeline}
\label{sec:exam_serve}

This subsection examines the serving latency of \textit{de facto} standard cloud serving and a vanilla fog deployment to investigate how much performance fog computing can promote.

\textbf{Methodology.}
The measurement targets a location-based recommendation application \cite{khanfor2020graph} that runs GCN inference on the SIoT dataset \cite{marche2020exploit} (dataset details in Table \ref{table:dataset} and \S\ref{sec:setup}).
The used graph includes 16216 devices from Spain as vertices with 146117 social connections, and each vertex attaches a 52-dimension feature that identifies its properties such as the device's type and brand.
Initially, the data are randomly divided into equal parts and assigned to 8 Raspberry Pis.
During the measurement, we launch the Pis to simultaneously send the respective graph data via 4G/5G/WiFi network, and then perform inference on the cloud/fog based on \textit{PyTorch Geometric} (\textit{PyG}) \cite{fey2019fast} once the complete graph is received.
The cloud server is an Aliyun instance (8vCPUs $\mid$ 32GB $\mid$ Tesla V100 GPU $\mid$ Ubuntu 16.04) located in the nearest region, and its geographical distance to the Pis is about 200km.
The fog cluster consists of six heterogeneous servers (specifications in \S\ref{sec:setup}) as computing nodes, all set on the same campus as the Pis.
In particular, for single-fog serving, we select the most powerful one to execute; for multi-fog serving, we apply the state-of-the-art technique in \cite{zheng2020distdgl} to place the input data among fog nodes and perform collaborative execution.
During the runtime, each node maintains a local graph, computes GNN layers, and exchanges vertices data with each other when needed.
The 4G/5G network employs commercial operator services, where the 5G network is provided by the 5G base stations surrounded nearby under the non-standalone (NSA) mode.

\begin{figure}[t]
  \centering
  \setlength{\abovecaptionskip}{0.1cm}
  \setlength{\belowcaptionskip}{-0.5cm}
  \includegraphics[width=0.9\linewidth]{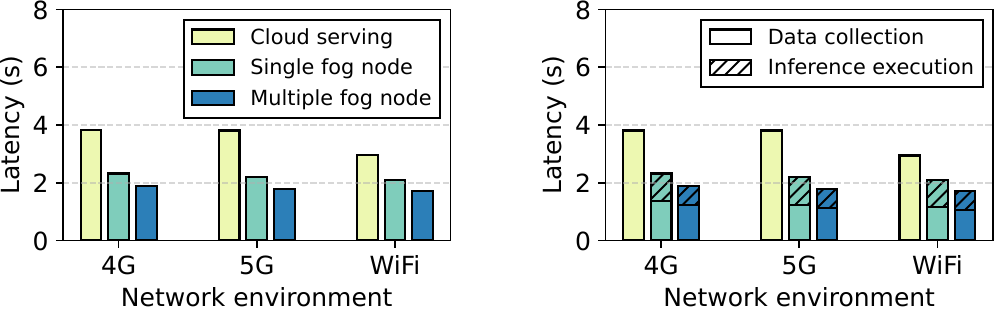}
  \caption{The latency measurements of three types of GNN serving (left) and their stage-wise breakdown (right).}
  \label{fig:motivation_latency}
\end{figure}

\begin{figure}[t]
  \centering
  \includegraphics[width=0.9\linewidth]{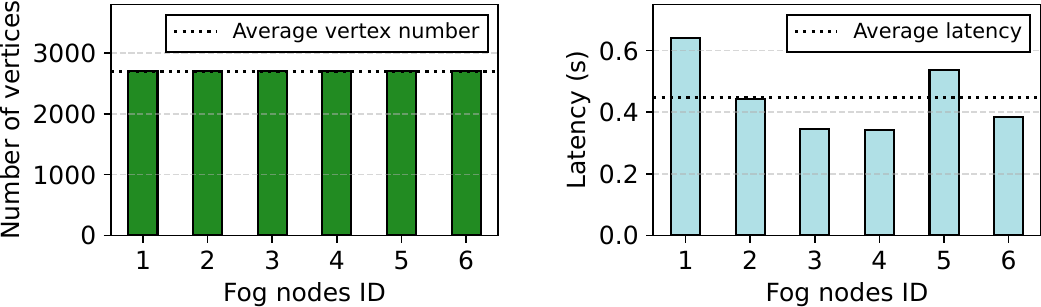}
  \caption{The number of vertices (left) and the inference execution latency (right) of each fog node in multi-fog serving.}
  \label{fig:motivation_load}
\end{figure}

\textbf{Measurements reported.}
Fig. \ref{fig:motivation_latency} (left) shows the serving latency of the cloud, single-fog and multi-fog mechanisms under different networking settings, and Fig. \ref{fig:motivation_latency} (right) breaks down stage-wise costs in terms of data collection and inference execution.
Regarding the load distribution in the multi-fog serving, Fig. \ref{fig:motivation_load} visualizes the number of assigned vertices and the execution latency of each fog node.

\textbf{Key observations.}
First, the fog approaches enjoy better performance than the cloud alternative, demonstrating its efficiency in vicinal serving.
Quantitatively, for 4G, 5G, and WiFi, the single-fog approach achieves 1.65$\times$, 1.73$\times$, and 1.40$\times$ speedups over the cloud serving, respectively.
The multi-fog counterpart attains even lower latency than the single-fog.
The weaker the networking condition is, the more superiority the fog serving reaps.

Second, we observe that the favour of fog serving is mainly contributed by communication.
As evidence, when switching from cloud to single-fog, the data collection latency can be reduced by 64\%, 67\%, and 61\% under 4G, 5G, and WiFi, respectively.
Such a similar degree of reductions implies the consistent advantages gained by the avoidance of remote Internet data transfer.
Surprisingly, multi-fog serving achieves lower costs in data collection than single-fog.
It is because employing more fog nodes provides more access points and therefore widens the bandwidth and relieves the networking contention.
Nonetheless, data collection still occupies $>$50\% costs in both fog approaches as in Fig. \ref{fig:motivation_latency} (right), suggesting that communication is yet the major spending factor in the serving pipeline.

Third, while the fog data collection significantly saves the overhead, its execution can dramatically compromise the benefit.
Nearly half of the costs are taken by execution in single-fog serving, while that in the cloud is $<$2\%.
Multi-fog serving alleviates that, but only reduces 33\% execution cost upon single-fog with five more fog nodes.
Such inferior performance indicates poor resource utilization, which comes from the gap between equally assigned data placement and heterogeneous computing resources.
This can be clearer from the measurements in Fig. \ref{fig:motivation_load}, where the existing data placement strategy merely yields an equilibrium in the number of assigned vertices but a severe imbalance in actual load distribution.

\subsection{Opportunities and Challenges with Fog Computing}

The advanced ability of GNNs has spread themselves to a wide range of end adoptions with real-time requirements.
Fog computing, as evident by the above realistic measurements, is revealed to be a potentially effective solution to address it, with both opportunities and challenges.

\textbf{Opportunities.}
By relocating execution to the approximate computing nodes close to the data sources, fog computing manages all data communication within a local network and thus avoids the unreliable and delay-significant Internet connections.
Such architectural wisdom, as proved by the above empirical measurements, is translated effectually into the reduction of communication time. 
In addition, the multi-node fog cluster provides more space for parallel GNN processing, which can further accelerate the serving pipeline.

\textbf{Challenges.}
In spite of the opportunities, simply adopting fog computing is not competent.
First, to effectively exploit the fog resources, it is imperative to accurately characterize the fog nodes' heterogeneity, decide the graph data placement, and orchestrate the data flow during the runtime.
Second, regarding the major contribution of transmission cost, a communication-friendly compression technique is desired for expedited graph data collection.
Third, to enable resilient serving in real-time, the system should be able to react dynamically to resource fluctuation.

\section{Fograph System Design} \label{sec:design}

As aforementioned, the objective is to fully unleash the architectural benefits of fog computing in rendering real-time GNN serving while adapting to fog nodes' heterogeneity and dynamic changes.
To this end, we propose Fograph, a distributed GNN inference system.
In what follows, we present Fograph modules following its workflow.

\subsection{Workflow and Design Overview} \label{sec:overview}

\begin{figure}[t]
  \centering
  \includegraphics[width=0.95\linewidth]{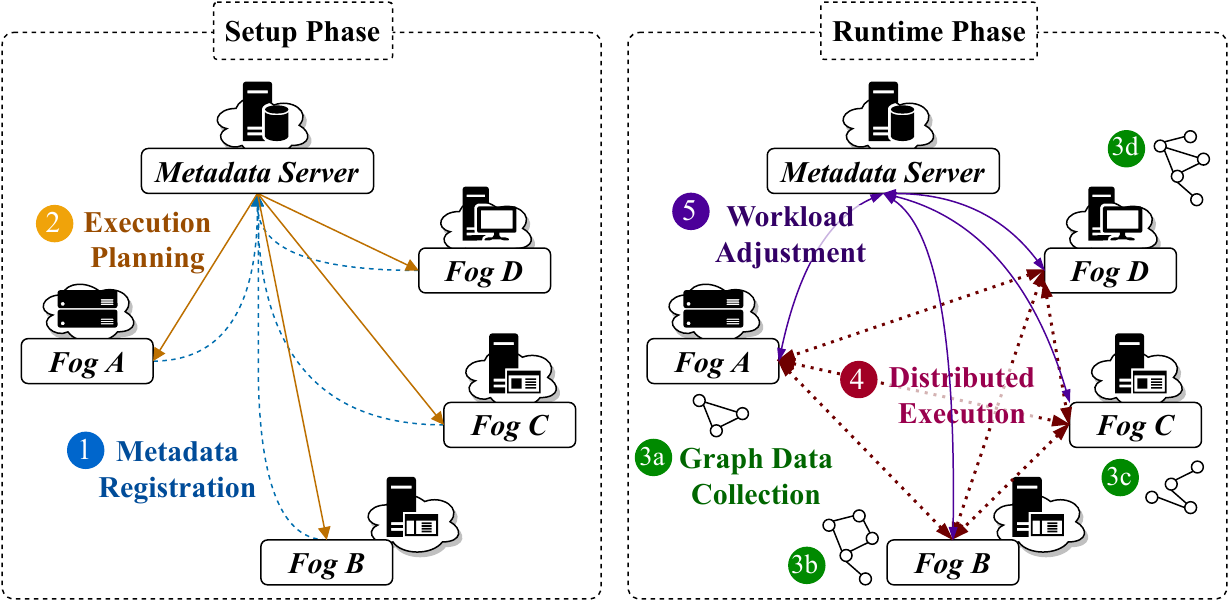}
  \caption{Fograph workflow overview. In the setup phase, Fograph first profiles and registers metadata, \textit{e.g.} model parameters and capability estimates, to the metadata server, and next decides a graph data placement among fog nodes through the execution planner. In the runtime phase, fog nodes collect graph data from devices, and perform collaborative execution for inference results. Simultaneously, Fograph monitors the load variation of fog nodes and dynamically adjusts the data placement to ensure resilient efficient serving.}
  \label{fig:workflow}
\end{figure}

Fig. \ref{fig:workflow} and Fig. \ref{fig:overview} show the high-level view of Fograph's workflow and system design, respectively, where the five modules in Fig. \ref{fig:overview} work for the five steps in Fig. \ref{fig:workflow} correspondingly.
In the setup phase, Fograph obtains a GNN model and uses the calibration dataset to sketch the computing capabilities of the heterogeneous fog nodes.
This is accomplished by the offline profiler (Fig. \ref{fig:overview} \ding{202}), which builds latency estimation models for predicting the GNN performance.
Moreover, it records the static properties of the historical input like the adjacency matrix, acting as the initial graph skeleton.
A dedicated fog node is selected as the metadata server that is responsible for registering metadata from all available fog nodes (\S\ref{sec:metadata}).
Next, Fograph's execution planner (\S\ref{sec:planning} \ding{203}) applies the latency estimates and judiciously schedules a graph data placement to match the heterogeneity among fog nodes, aiming at maximum parallelization with effective load balance guarantee.

In the runtime phase, the participated fog nodes individually collect their assigned data partitions in light of the execution plan.
To speed up device-to-fog data transfer, a novel GNN-specific compression technique is employed to exploit the features sparsity and GNN's resilience to progressively reduce data uploading costs (\S\ref{sec:compression} \ding{204}).
Once input graph data completely arrive, Fograph's runtime orchestrates distributed inference execution, handling all data exchange between fog nodes (\S\ref{sec:execution} \ding{205}).
Simultaneously, each fog node's online profiler monitors its resident execution across inferences, as well as the runtime conditions, and updates the offline performance profile, periodically transferring to the metadata server for execution plan refinement (\S\ref{sec:execution} \ding{206}). 
In this way, the system can adapt to dynamic-changing environments,
reconfigure its execution and maintain real-time serving.

\subsection{Metadata Acquisition and Registration}
\label{sec:metadata}

The aim of \textit{metadata registration} (Fig. \ref{fig:workflow} \ding{202}) is to readily provision fundamental serving configurations and sensibly characterize the heterogeneity of fog nodes, providing necessary materials for the subsequent execution planning.
To achieve that, we design a dynamic profiler, operating across the setup phase and the runtime phase.

\begin{figure}[t]
  \centering
  \includegraphics[width=0.9\linewidth]{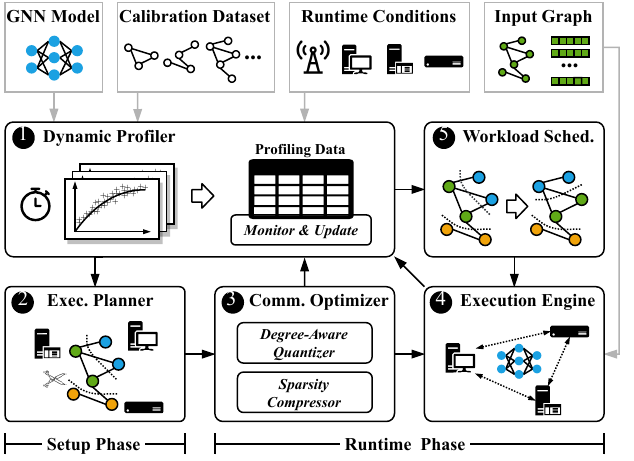}
  \caption{Fograph design overview, where the five modules correspondingly work for the five steps in Fig. \ref{fig:workflow}.}
  \label{fig:overview}
\end{figure}

\textbf{Setup phase.}
Before deployment, the offline profiler (Fig. \ref{fig:overview} \ding{202}) performs two kinds of measurements, device-independent and device-specific.
The former focuses on the static configurations stated by service providers.
Concretely, it comprises 1) available bandwidth of fogs, 2) the employed GNN model (trained in advance), and 3) the invariant metrics of the input graph.
Here we identify the invariance as the adjacency matrix/list that depicts the graph topology, and the size of a feature vector, which is determined once a given GNN model is trained.
For instance, in smart transportation applications \cite{guo2019attention, yu2017spatio, ye2020build}, we can interpret them as the traffic monitoring sensors' logical topology (\textit{e.g.} sensors as vertices and roads as edges) and the form of sensory records, both of which are known before runtime.
These parameters are independent of the running platforms and thus can be profiled only once for a given model.

For the latter, we intend to establish latency estimation models that are specific to each fog node, targeting quantifying their heterogeneous computing capability.
The performance of computing GNN inference, however, relies heavily on the fundamental settings such as the underlying hardware and the used DL framework, where trivial estimations based on static configurations are rough and unfaithful.
Therefore, to build performance models in a precious granularity, we employ a proxy-guided profiling process:
First, we construct a calibration set by uniformly sampling subgraphs of varying cardinality from the initial graph.
The cardinality, defined as $\langle c \rangle=\langle |\mathcal{V}|, |\mathcal{N}_{\mathcal{V}}| \rangle$, shapes a subgraph's size from a GNN perspective with the number of vertices it includes and their one-hop neighbors.
To reserve the actual degree distribution, for each cardinality axis we collect a group of 20 samples.
Next, we measure the average execution latency for each fog node by passing the GNN through the calibration set, and build regression-based latency estimation models $\omega$, \textit{e.g.} linear regression model in Eq. (\ref{eq:regression}) where $\beta$ and $\varepsilon$ are regression parameters.
\begin{align}
    \text{latency} = \omega(\langle c \rangle) = \beta \cdot \langle |\mathcal{V}|, |\mathcal{N}_{\mathcal{V}}| \rangle + \varepsilon. \label{eq:regression}
\end{align}

\textbf{Runtime phase.}
At run time, the profiler keeps tracking the execution time of each fog node to update the offline estimates and derives the balance indicators in order to gauge the global performance.
To keep the profiler lightweight, instead of adopting a more accurate but prohibitively costly estimator, we employ a two-step linear estimation to predict the inference latency on the fly.
In the first step, the profiler measures the actual execution time of the local $c$-cardinality graph, denoted by $T^{\text{real}}_{\langle c \rangle}$, during each inference.
Next, it calculates a load factor $\eta$ as the ratio
between the actual time and the offline latency estimate of cardinality $c$, \textit{i.e.} $\eta = \frac{T^{\text{real}}_{\langle c \rangle}}{\omega(\langle c \rangle)}$.
As the second step, the profiler treats the load factor as a reflection on the present load level, and uses it to predict the latency of all other cardinalities.
Thus, the latency of a different cardinality $c'$ is estimated as $\eta \cdot \omega(\langle c' \rangle)$.

\subsection{Inference Execution Planning}
\label{sec:planning}

Given a GNN model, Fograph exploits data parallelism to distribute the inference workloads over multiple fog nodes, where input data needs to be divided and distributed.
To attain high-performing serving, an \textit{inference execution planner} (IEP, Fig. \ref{fig:overview} \ding{203}) is developed to schedule data placement ahead of runtime.

\textbf{Problem formulation.} Let $\mathcal{G} = (\mathcal{V}, \mathcal{E})$ define a GNN input graph, where $\mathcal{V}$ and $\mathcal{E}$ are the set of vertices and edges, respectively.
Suppose a set $\mathcal{F}$ of $n$ fog nodes are available in serving, denoted by $\langle f_1, f_2, \cdots, f_n \rangle$.
Each vertex $v_i \in \mathcal{V}$ is a data source point (\textit{e.g.} a sensor), and its placement to a certain fog $f_j$ is specified by a binary variable $x_{ij} \in \{0, 1\}$.
While a fog admits multiple vertices, a vertex can only be placed to exactly one fog, \textit{i.e.},
\begin{align}
     \textstyle\sum_j x_{ij} = 1, \ \forall v_i \in \mathcal{V}. \label{eq:variable_constraint}
\end{align}

To reckon the cost of a fog's data collection process, we should tally the transmission latency of all vertices placed to it, as in Eq. (\ref{eq:data_collection}), where $\varphi$ is the data size of a single vertex's feature vector and $b_{j}$ indicates the $f_j$'s available bandwidth for serving.
\begin{align}
    % t^{\text{colle}}_{j} &= \max_i(\frac{x_{ij}d}{b_{ij}}), \ \forall f_j \in \mathcal{F}. 
    t^{\text{colle}}_{j} &= \frac{\sum_i x_{ij} \varphi}{b_{j}}, \ \forall f_j \in \mathcal{F}. 
    \label{eq:data_collection}
\end{align}

To calculate the inference execution latency on $f_j$, we summarize its placed vertices as a subgraph $\textstyle\bigcup_{i} x_{ij} v_i$, and estimate its computing time through the performance model $\omega(\cdot)$ from metadata.
Besides, for the complete execution runtime, there are additionally $K$ synchronizations for cross-fog data exchange through a $K$-layer GNN, due to GNN's neighbor aggregation mechanism as discussed in \S\ref{sec:gnn}.
Assuming the cost of a synchronization is $\delta$, we append $K\delta$ to complement the total execution cost, as in Eq. (\ref{eq:execution}).
\begin{align}
    t^{\text{exec}}_{j} &= \omega_j(\textstyle\bigcup_{i} x_{ij} v_i) + K\delta, \ \forall f_j \in \mathcal{F}. \label{eq:execution}
\end{align}

Putting both data collection and inference execution together, the objective of IEP is to find an efficient data placement strategy $\pi = \{x_{ij}|\forall v_i \in \mathcal{V}, \forall f_j \in \mathcal{F}\}$ such that the latency of the complete serving pipeline is minimized, formally formulated in problem $\mathcal{P}$:
\begin{align}
    \mathcal{P}: \text{min} & \ \textstyle\text{max}_j(t^{\text{colle}}_{j} + t^{\text{exec}}_{j}), \label{eq:objective} \\
    \text{s.t.} & \ (\ref{eq:variable_constraint}), (\ref{eq:data_collection}), (\ref{eq:execution}). \notag
\end{align}
\begin{theorem}
    \label{theorem:NP_hard}
    The IEP problem $\mathcal{P}$ is NP-hard when the number of fog nodes $n \geq 2$.
\end{theorem}

The quality of this data placement matters in that 1) uneven assignment usually catalyzes the straggler effect \cite{cipar2013solving} and 2) skew load distribution commonly accompanies communication bottlenecks, and either of them can largely slow down the parallel performance as measured in \S\ref{sec:exam_serve}.
However, to yield an optimal solution of $\mathcal{P}$ is intractable when the number of fogs $n \geq 2$, due to its NP-hardness stated in Theorem \ref{theorem:NP_hard} (proof in Appendix A).
Unfortunately, the integration of the unique computation pattern enforced by GNN workload and the inherent heterogeneity of fog nodes makes existing data placement techniques hardly be applied.

\textbf{IEP data placement.}
To enable real-time GNN serving among fogs, we alternatively leverage heuristics to make efficient optimization.
Specifically, we capitalize two insights in IEP: 
1) Co-locating vertices with connections to a mutual data partition can not only save its computing time $\omega_j(\textstyle\bigcup_{i} x_{ij} v_i)$ on a single fog but alleviate the synchronization burden $K\delta$ in Eq. (\ref{eq:execution}).
This knob attributes to GNN's neighbor aggregation mechanism, where computing inference over a data partition is essentially operating on a neighbor-augmented graph upon the vertices it contains.
Maximizing the vertices' locality within a data partition can thus significantly decrease the number of neighbors $|\mathcal{N}_{\mathcal{V}}|$, so that the computing time (refer to Eq. (\ref{eq:regression})) and synchronization costs (for pulling neighbors from other fogs) are lessened simultaneously.
2) In light of the parallel nature of $\mathcal{P}$, a serving-oriented load balance with regards to heterogeneous computing capability and diverse bandwidth can profitably promote the holistic performance.
Particularly, the costs of data collection $t^{\text{colle}}_j$ and inference execution $t^{\text{exec}}_j$ should be jointly considered to maximize the utilization of available resources.
Motivated by these two insights, we tackle $\mathcal{P}$ via a two-step optimization that first preprocesses the input graph to generate locality-maximized partitions, and next maps them to fog nodes accounting for both computation and communication resources.

\begin{figure}[t]
  \centering
  \includegraphics[width=\linewidth]{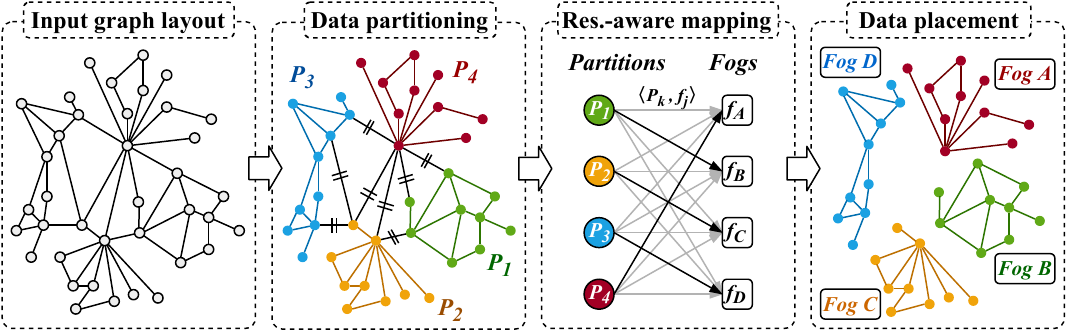}
  \caption{Illustration of Fograph's data placement algorithm, which primarily comprises of a graph partitioning step and a resource-aware matching step.}
  \label{fig:algorithm}
\end{figure}

Fig. \ref{fig:algorithm} depicts its overall flow in IEP and Algorithm \ref{algo:data_placement} shows the pseudocode.
First, we intend to generate data partitions over the input graph, aiming at both internal vertex locality and load balancing.
Instead of searching by brute force, we remark that this task is partially related to Balanced Graph Partitioning (BGP), a family of problems that have been extensively studied \cite{pacaci2019experimental, abbas2018streaming}.
Particularly, maximizing the partitions' internal locality can be conversely interpreted to a minimization on inter-partition connections, \textit{i.e.} edge-cuts.
Therefore, as an initialization, we employ BGP solvers and attain a group of $n$ min-cut data partitions, where $n$ is the number of fogs (Line 2).
These partitions, however, are merely statistically balanced in vertices' number rather than actual GNN workload, which may still induce uneven load distribution (as discussed in \S\ref{sec:exam_serve}).
To bridge this gap, in the second step, we build a resource-aware mapping between partitions and fogs.
Specifically, a bipartite graph $\mathcal{B}$ is defined as in Fig. \ref{fig:algorithm} (\textit{Res.-aware mapping} box), with partitions $\langle P_1, P_2, \cdots, P_n \rangle$ and fogs $\langle f_1, f_2, \cdots, f_n \rangle$ in separate columns.
We associate every partition-fog pair with an edge, weighted by a compositive cost $\langle P_k, f_j \rangle$ of uploading and executing partition $P_k$ on fog $f_j$:
\begin{align}
    \langle P_k, f_j \rangle = \textstyle\frac{|P_k| \varphi}{b_j} + \omega_j(P_k) + K\delta, \ k,j \in \{1,2,\cdots,n\}. \label{eq:bipartite_weight}
\end{align}

\begin{algorithm}[t]
\caption{IEP resource-aware data placement}
\label{algo:data_placement}
\KwIn{
    Input data graph $\mathcal{G}$ \\
    \qquad \quad \ Performance estimation model $\omega$ \\
    % \qquad \quad \ Available bandwidth $\langle b_1, b_2, \cdots, b_n \rangle$\\w
}
\KwOut{
    Data placement $\pi$
}

\textit{/* - - - Step 1: data partitioning - - - */}

$\langle P_1, P_2, \cdots, P_n  \rangle \gets \text{BGP}(\mathcal{G}) $ 

\textit{/* - - - Step 2: partition-fog mapping - - - */}

Construct a partition-fog bipartite graph $\mathcal{B}$ with $\langle P_1, P_2, \cdots, P_n \rangle$ and $\langle f_1, f_2, \cdots, f_n \rangle$

Assign weights to edges in $\mathcal{B}$ according to Eq. (\ref{eq:bipartite_weight})

Put all edge weights to a priority queue $\mathcal{Q}$

\While{$\mathcal{Q} \neq \emptyset$}{
    $\tau \gets \mathcal{Q}.\texttt{dequeue()}$
    
    $\mathcal{B}' \gets \mathcal{B}$
    
    Filter edges in $\mathcal{B}'$ with weight threshold $\tau$
    
    $M \gets \text{Hungarian}(\mathcal{B}')$
    
    \eIf{$M$ is a perfect mapping}{
        $M^* \gets M$
    }{
        \textbf{break}
    }
    
}

Construct $\pi$ according to $M^*$

\KwRet{$\pi$}
\end{algorithm}

Yet to find a mapping in $\mathcal{B}$ that satisfies $\mathcal{P}$'s min-max objective has a huge decision space of $n!$, and differs from the traditional bipartite matching problem of maximum weighted sum \cite{lovasz2009matching}.
However, we observe that it is a variant of \textit{Linear Bottleneck Assignment Problem (LBAP)} \cite{burkard1999linear}, and we can apply a threshold-based algorithm to solve an optimal mapping.
Specifically, it first instantiates a priority queue $\mathcal{Q}$ to accommodate all edge weights in $\mathcal{B}$, and next successively inspects every element in iterations.
For each iteration, it dequeues the front element in $\mathcal{Q}$, the maximum weight in the queue, as the weight threshold $\tau$, and filters edges in $\mathcal{B}$ that have a weight smaller than $\tau$ to construct an auxiliary bipartite graph $\mathcal{B}'$ (Line 8-10).
Applying the Hungarian algorithm \cite{munkres1957algorithms}, we obtain a mapping $M$ in $\mathcal{B}'$ and check whether it is a perfect matching towards the original bipartite graph $\mathcal{B}$.
If succeed, we record the obtained mapping in $M^*$, and move forward to another iteration for new attempts with lower thresholds.
Otherwise, there is no perfect matching anymore in filtered bipartite graphs since the remaining to-be-examined thresholds in $\mathcal{Q}$ will be smaller; the obtained mapping $M^*$ is thus the expected result that minimizes the maximum weight in $\mathcal{B}$ and the iteration consequently terminates.
Finally, the algorithm ends by returning the $M^*$'s corresponding data placement $\pi$.

\begin{figure}[t]
  \centering
  \includegraphics[width=0.96\linewidth]{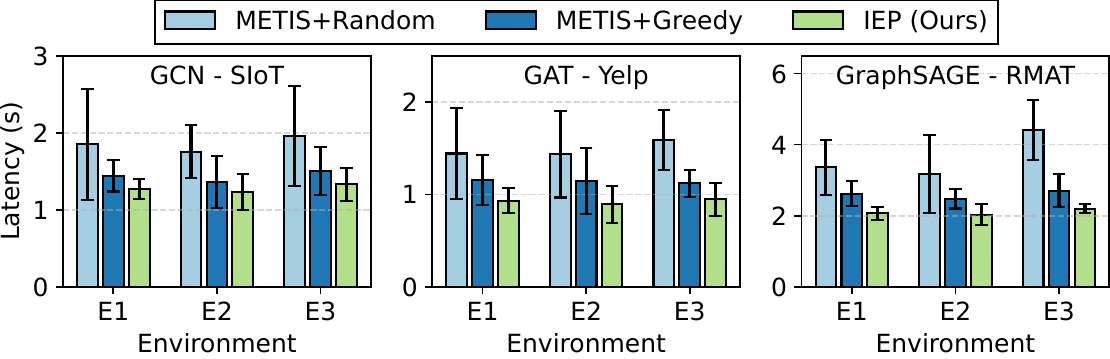}
  \caption{
  Performance comparison of IEP against straw-man approaches in three environments. Given the three types of fog nodes listed in Table \ref{table:hardware}, their hardware and network conditions can be represented as follows. E1: \{1$\times$\textit{A}, 4$\times$\textit{B}, 1$\times$\textit{C}, 4G\}; E2: \{1$\times$\textit{A}, 4$\times$\textit{B}, 1$\times$\textit{C}, 5G\}; E3: \{1$\times$\textit{A}, 2$\times$\textit{B}, 1$\times$\textit{C}, WiFi\}. 
  }
  \label{fig:iep_ablate}
\end{figure}

\textbf{Discussion.}
The expense of IEP's first step mainly relies on the selected BGP solver, and Fograph allows for altering appropriate solvers to adapt to specific graphs and reduce the overhead.
The second step takes $O(n^2)$ iterations for threshold descending, according to the $O(n^2)$ length of the priority queue from the bipartite graph $\mathcal{B}$ with $n$ partitions and $n$ fogs.
However, we can use binary search to further expedite the threshold searching, which can significantly decrease overall iterations to $O(\log n)$.
As each iteration's Hungarian algorithm invocation requires $O(n^3)$, the second step of IEP takes a total time complexity of $O(n^3 \log n)$.
We note that such a complexity is affordable since the number of available fog nodes in real deployment is usually small (\textit{e.g.} $<$100).
Besides, the scheduling overhead of IEP data placement is an upfront cost before runtime and it can be amortized across multiple inferences.
In our implementation, we apply the widely-adopted \textit{METIS} \cite{karypis1997metis} as the BGP solver and binary search for the mapping step, which spends only seconds for SIoT in total.

To verify the effectiveness of the proposed IEP algorithm, we examine its performance against two comparative straw-man approaches: 1) METIS+Random, a trivial version that first invokes METIS for balanced partitions and next assigns them to arbitrary fog nodes, and 2) METIS+Greedy, which takes a greedy heuristic in IEP's partition-fog mapping procedure, \textit{i.e.} iteratively finds fogs for partitions such that their edge weight $\langle P_k, f_j \rangle$ is minimized.
Fig. \ref{fig:iep_ablate} depicts the results in three heterogeneous environments, where we observe that IEP always surpasses baselines with lower serving latency, demonstrating the superiority of our resource-aware algorithm design. 
Specifically, the updated IEP algorithm outperforms METIS+Greedy with an average latency reduction of 10.9\%, 19.1\%, and 19.5\% for three different model configurations, respectively.

\subsection{Degree-Aware Compressed Data Collection}
\label{sec:compression}

According to IEP's data placement, each fog collects the input data individually in the runtime phase (Fig. \ref{fig:workflow} \ding{204}).
As discussed in \S\ref{sec:exam_serve}, the considerable costs of uploading graph data stress its significance towards real-time serving.
To alleviate this bottleneck, Fograph integrates a \textit{communication optimizer} (CO, Fig. \ref{fig:overview} \ding{204}), operating in two steps.

\textbf{Degree-aware quantization} (DAQ). 
In the first step, we exploit the resilience of GNNs to low-precision representation and lower the data precision in a differentiated way with the topological information.
Concretely, we use each vertex's degree as a knob to modulate the quantization intensity on its feature vector.
The rationale behind is that a vertex with a higher degree assimilates more abundant information from its neighbors, and is more robust to low bitwidths since its quantization errors can be smooth through successive aggregations \cite{feng2020sgquant}.

\begin{figure}[t]
  \centering
  \includegraphics[width=0.9\linewidth]{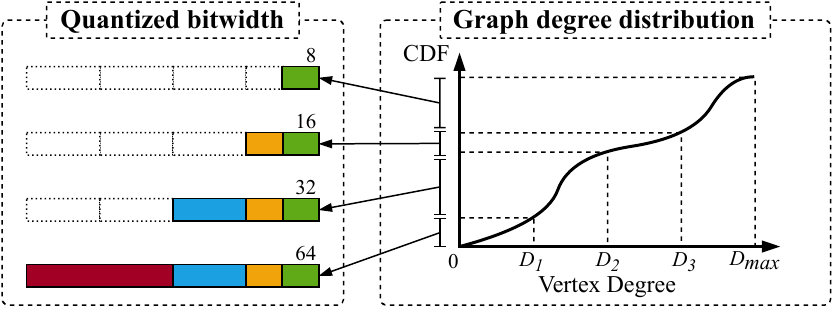}
  \caption{Illustration of Fograph's degree-aware quantization. Each feature vector is quantized to a targeted bitwidth according to the degree of the vertex that it associates.}
  \label{fig:quantization}
\end{figure}

In detail, DAQ maintains a triplet $\langle D_1, D_2, D_3 \rangle$ to divide the vertices' degrees into four intervals $[0, D_1)$, $[D_1, D_2)$, $[D_2, D_3)$ and $[D_3, +\infty)$, and assigns respective quantization bits of $\langle q_0, q_1, q_2, q_3 \rangle$ to each.
Next, for each vertex, we check its degree to obtain a target bitwidth according to the interval it lies in, and implement a linear quantization.
In Fograph, we reckon up four equal-length intervals based on the input graph's degree distribution and instantiate the quantized bits as $\langle 64, 32, 16, 8 \rangle$ by default, as illustrated in Fig. \ref{fig:quantization}.
However, it should be noted that $\langle D_1, D_2, D_3 \rangle$ and $\langle q_0, q_1, q_2, q_3 \rangle$ are tunable to accommodate specific graph topology and customized accuracy-latency preference.
The exploration of these configurations is left for future work.

\begin{theorem}
    \label{theorem:compression_ratio}
    Given the input feature's bitwidth as $Q$, DAQ with configurations $\langle D_1, D_2, D_3 \rangle$ and $\langle q_0, q_1, q_2, q_3 \rangle$ renders a compression ratio of $\frac{1}{Q}[q_3 - \sum_i F_D(D_i)(q_i - q_{i-1})], i\in\{1,2,3\}$,  where $F_D(\cdot)$ is the cumulative distribution function of the graph degree distribution.
\end{theorem}

Theorem \ref{theorem:compression_ratio} theoretically gives the compression ratio of DAQ, with proof in Appendix B.
By discriminatively reducing the bitwidths of feature vectors, DAQ allows the transferred size to be substantially lower without significant impact on the inference accuracy ($<$1\% drop in our experiments).
Our scheme differs from both 1) uniform quantization \cite{feng2020sgquant}, which ignores the vertices' different quantization sensitivity and degrades accuracy, and 2) all-layers quantization that demands complicated techniques \cite{tailor2020degree}, such as quantization-aware training, to fine-tune the model prediction performance.

\textbf{Sparsity elimination.}
The second step exploits the observation that feature vectors are amenable to compression. 
A major fraction of feature vectors are sparse and highly compressible due to their encoding mechanism, and the sparsity is further magnified by precision reduction in the above quantization step. 
Hence, we compress the sparsity using the \textit{LZ4} compressor \cite{LZ4_compression} with bit shuffling.

\textbf{Deployment of CO.}
The lifecycle of CO comprises procedures of packing and unpacking, where the former is deployed at end devices that contribute data sources and the latter is installed on fog nodes.
While its cost can be amortized by the communication savings, we develop several optimizations to further reduce the overhead.
For packing, we mitigate the burden on end devices by
1) pre-calculating the targeted quantization bitwidth before deployment and using it through the runtime (as long as the graph topology is unchanged), and
2) quantizing the feature vector's elements in parallel for additional acceleration.
Since each device uploads its local data individually, the data packing process is naturally parallelized from a global view and the cost is apportioned.
For the fog side, the received data are first decompressed and then dequantized back to the original bitwidth before inference.
Further, we launch a separated thread for the unpacking procedure to pipeline data recovering and inference execution.

\subsection{Distributed Execution Runtime}
\label{sec:execution}

With the generated execution plan, Fograph's runtime \textit{execution engine} (Fig. \ref{fig:overview} \ding{205}) orchestrates distributed GNN inference upon multiple nodes.
Specifically, when an inference query is launched, fog nodes will collect the necessary data from nearby sensory devices as per the data placement policy, and then collaborate to conduct the GNN execution. 
For each GNN layer execution, cross-fog data exchanges will be carried out when necessary, \textit{i.e.} neighboring vertices' data belong to different data partitions.
Next, inference functions (\texttt{Aggregate} and \texttt{Update}) are invoked by the fog nodes to compute the layer over the data partitions in parallel.
Repeating the above process for all layers completes the whole execution and produces expected embeddings.
Note that the model has been loaded and stays resident throughout the runtime so that it can be called immediately on demand.
To adapt to dynamic fluctuation of resources, \textit{e.g.} background load and bandwidth changes, the metadata server periodically aggregates the metadata from fog nodes, replays IEP to yield an updated data placement, and deploys to fogs at system idle time.

The iterative layer processing is implemented with the Bulk Synchronous Parallel (BSP) model \cite{valiant1990bridging}, where a synchronization step is triggered when data exchange is needed.
Although the total synchronization times depend on the number of GNN layers, which is usually very small (\textit{e.g.} GCN typically stacks two or three layers), we apply the following optimizations for further acceleration.
First, the adjacency matrix of each data partition can be constructed prior to the execution as long as the data placement is determined, in order to lower the occupancy of runtime.
Second, the synchronization is run as a separated thread to enable the pipelining of data preparation and inference execution.
Third, we wrap the execution on top of the mature framework \textit{PyG} \cite{fey2019fast}, which allows to directly benefit from all existing kernel-level optimizations.

\subsection{Adaptive Workload Scheduling} \label{sec:adapter}

Even with the offline best graph data placement, the distributed inference performance can reach suboptimum due to the fluctuation of computing resource, \textit{e.g.} caused by machine load variation.
This can be relevant considering that fog nodes usually run versatile services simultaneously.
To this end, the \textit{adaptive workload scheduler} (Fig. \ref{fig:overview} \ding{206}) is developed to refine the data placement tailored to dynamic load levels.
Unlike the offline IEP that makes meticulous yet expensive optimization, the adaptive scheduler works online and should keep agile and agnostic to the inference.
Therefore, we employ a lightweight adjustment method and a dual-mode regulation to adapt the workload distribution.

\begin{figure}[t]
    \centering
    \subfigure[Before diffusion.]{
        \begin{minipage}[t]{0.42\linewidth}
        \centering
        \includegraphics[width=\linewidth]{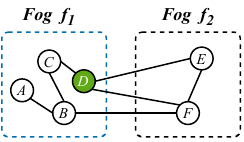}
        \label{fig:diff_before}
        \end{minipage}
    }
    \quad
    \subfigure[After diffusion.]{
        \begin{minipage}[t]{0.42\linewidth}
        \centering
        \includegraphics[width=\linewidth]{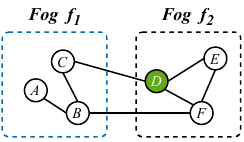}
        \label{fig:diff_after}
        \end{minipage}
    }
    \caption{An instance of diffusion-based graph data placement adjustment, where vertex $D$ is selected for migration from the overloaded fog node $f_1$ to fog node $f_2$.
    }
    \label{fig:diffusion}
\end{figure}

\textbf{Load balance indicator.} 
To reflect how skew the load distribution is, we first define a indicator $\mu_j$ for each node $f_j$ by inspecting the fraction between its actual execution time $T^{\text{real}}_j$ and the mean value of all fogs' measurements $\frac{1}{n}\sum_k T^{\text{real}}_k$:
\begin{align}
\mu_j = \frac{T^{\text{real}}_j}{\frac{1}{n}\sum_k T^{\text{real}}_k}, \quad \forall f_j \in \mathcal{F}, \label{eq:indicator}
\end{align}
where $T^{\text{real}}_j$ is the real measured execution time of $f_j$ in the last time interval, and is obtained from the online profiler.
We further introduce a slackness factor $\lambda$ to tune the imbalance tolerance.
If there is a node such that $\mu_j > \lambda$, it implies this node breaks the imbalance tolerance $\lambda$ and is supposed to suffer from a high background load.
Note that $\lambda$ is obliged to be larger than 1.
$\lambda = 1$ represents that exact balance is required whereas $\lambda > 1$ relaxes the balanced constraints.
Additionally, we count the number of overloaded nodes, denoted as $n^{+}$, to gauge the global load skewness and use it for the next configuration.

\textbf{Diffusion-based adjustment.}
The diffusion method aims at amending the graph data placement to align with the load level at a low cost.
With the latest profiling data, it first selects two partitioned sets of vertices with the highest and lowest execution time, and then progressively migrates the vertices from the overloaded set to the underloaded set until an estimated local balance is achieved.
For each migration, the boundary vertex that shares the most neighbors with the other side is picked.
As an example, Fig. \ref{fig:diffusion} illustrates this diffusion process across fog nodes $f_1$ and $f_2$, where they are supposed to be moderate and weak in computing capability, respectively.
Without external load burdens, a graph data placement is decided initially, separating four/two vertices as in Fig. \ref{fig:diff_before}.
Assuming a load increase abruptly happens in $f_1$ such that the two nodes' capabilities are currently on a par, the diffusion is then applied to migrate vertices for balancing their workload.
The number of to-be-migrated vertices is 1, and the adjustment will choose vertex \textit{D} as the moving candidate since it connects the most edge-cuts across subgraph in $f_1$ and $f_2$.
This consequently results in an updated balanced layout in Fig. \ref{fig:diff_after}.
In a data placement with multiple partitions, the above pairwise diffusion process continues for all unevenly-loaded nodes until the overall estimated performance satisfies the imbalance tolerance $\lambda$.

\begin{algorithm}[t]
\caption{Flow of adaptive workload scheduler}
\label{algo:adapter}
\KwIn{
    Current graph data placement $\pi$ \\
    \qquad \quad \ Load imbalance tolerance $\lambda$ \\
    \qquad \quad \ Skewness threshold $\theta$ \\
    \qquad \quad \ Performance estimation model $\omega$ \\
}
\KwOut{
    Updated graph data placement $\pi'$
}

$ \omega' \gets \text{UpdateTimings}(\omega, \langle T^{\text{real}}_1, \cdots, T^{\text{real}}_n \rangle)$ 

$\langle \mu_1, \cdots, \mu_n \rangle \gets \text{CalculateSkew}(\langle T^{\text{real}}_1, \cdots, T^{\text{real}}_n \rangle)$

\eIf{ $\exists f_j \in \mathcal{F}, \mu_j > \lambda$ }{
    
    $n^{+} \gets n_{\{\mu_i > \lambda\}}$
    
    \eIf{ $n^{+}/n \leq \theta$ }{ 
        \textit{/* - - - Lightweight adjustment - - - */}

        $\pi' \gets \text{DiffAdjust}(\pi, \omega', \langle \mu_1, \cdots, \mu_n \rangle ) $ 
    }{
        \textit{/* - - - Global rescheduling - - - */}
        
        $\pi' \gets \text{IEP}(\mathcal{G}, \omega') $ 
    }
}
{
    $\pi' \gets \pi$
}
\KwRet{$\pi'$}
\end{algorithm}

We refer to this method as \textit{diffusion} in that the flow of vertices continuously moves from the overloaded regions to the underloaded regions.
This method is lightweight since it only operates on a small part of the graph and migrates a few vertices.
Yet it is effective as it consistently makes incremental improvements on the graph layout.
However, when the load distribution is dramatically skew, this property also comes at a high cost.
Therefore, we introduce a dual-mode scheduler to integrate the lightweight adjustment with the global partitioning.

\textbf{Dual-mode scheduler.}
The workload scheduler considers dual regulations, the lightweight diffusion-based adjustment discussed above and the heavy global partitioning that invokes the offline IEP.
Algorithm \ref{algo:adapter} presents the scheduler's processing flow.
As a first step, the scheduler uses recorded execution time to update the performance estimation models and calculate the skewness indicators (Line 1-2). 
If there is a node such that $\mu_j > \lambda$, it implies the imbalance tolerance $\lambda$ is violated.
This subsequently triggers adjustments on the existing graph layout.
To decide which mode to be applied, we count the number of overloaded nodes, denoted as $n^{+}$, and compare it with a user-specific skewness threshold $\theta$ (Line 4-5).
$\theta$ is a positive decimal and is set 0.5 by default.
If $|n^{+}| / n \leq \theta$, it means that the skewness is still tolerable and the lightweight diffusion is applied.
Otherwise, the percentage of the overloaded nodes exceeds $\theta$ and we pass the entire graph $\mathcal{G}$ and the updated performance estimates $\omega'$ to IEP to yield a new data placement.
Note that all the layout modifications are first operated virtually in Algorithm \ref{algo:adapter} and will be executed physically when the final result is determined and the system is in idle period.

\section{Evaluation}
\label{sec:evaluation}

This section presents the effectiveness of Fograph in significantly improving the performance of GNN inference serving by examining its core components and comparing with the currently standard cloud implementations and straw-man fog inference systems.

\subsection{Experimental Setup}
\label{sec:setup}

\textbf{Prototype and methodology.}
We implement Fograph prototype with three types of computing nodes and their specifications are listed in Table \ref{table:hardware}.
We label them as type \textit{A}, \textit{B}, and \textit{C}, respectively representing fog nodes with weak, moderate, and powerful capabilities.
Such a category depends on their processors' power and available memory: Type-\textit{C} fogs equip the highest computing power with the largest memory space, whereas Type \textit{A} is on the opposite side and Type \textit{B} is in between them.
Although fogs of Type \textit{A} and \textit{B} own the same processor, the former performs much poorer than the latter with the used SIoT and Yelp datasets, reporting an average of 37.8\% longer inference latency with GCN.
Fograph is built on top of \textit{PyG} \cite{fey2019fast}, though its design is agnostic to the backend and can be conveniently switched to other engines like \textit{DGL} \cite{wang2019deep}.
We compare Fograph with the \textit{de-facto} standard cloud serving and a straw-man multi-fog deployment (multi-fog serving without any proposed optimization), all running the same workload.
The configurations of cloud and the emulation of distributed graph data uploading follow the methodology in \S\ref{sec:exam_serve}.
To ensure a fair comparison, the straw-man fog approach adopts the data placement strategy in state-of-the-art distributed GNN processing \cite{zheng2020distdgl} that first calls \textit{METIS} \cite{karypis1997metis} to partition the data graph and next map them to fog nodes stochastically.
According to the placement, the fog approach's runtime directly collects graph data without communication optimization, and launches collaborative inference upon the same distributed framework as Fograph.
Without loss of generality, Fograph selects an arbitrary fog node as the metadata server among the used ones.
We measure the end-to-end latency from data collection to GNN inference completion.
All the background loads are disabled during the runtime and each benchmark is evaluated 50 times to calculate the average results.

\textbf{GNN models and datasets.}
Four GNN models are employed: GCN \cite{kipf2016semi}, GAT \cite{velivckovic2017graph}, GraphSAGE \cite{hamilton2017inductive} and ASTGCN \cite{guo2019attention}. 
The first three are representative GNN models that have been widely used across GNN-based services \cite{wu2020comprehensive}, and their formulas have been listed in Table \ref{table:models}.
ASTGCN is a spatial-temporal model specific for traffic flow forecasting.
All the models are implemented using the instances from \textit{PyG} model zoo \cite{pyg_github} or the original code from the model authors \cite{astgcn_github}, and are trained prior to deployment.

We evaluate Fograph on three real-world datasets and their statistics are listed in Table \ref{table:dataset}.
We select them since they have been used in existing literature \cite{zhong2020hybrid, dou2020enhancing, guo2019attention}, and are yet the largest publicly available ones at the time of evaluation, regrading the real-world IoT-driven smart applications discussed in \S\ref{sec:gnn}.
We also synthesize much larger graphs of different scales, namely the RMAT series in Table \ref{table:dataset}, upon the above real-world datasets to examine system scalability.
The detailed description of the datasets is presented in Appendix C and Appendix D.

\begin{table}[t]
    \caption{Specifications of used fog nodes.}
    \label{table:hardware}
    \centering
    \setlength{\belowcaptionskip}{-0.5cm}
    \begin{tabular}{|c|c|c|c|} 
        \hline
        \textbf{Type} & \textbf{Processor} & \textbf{Memory} &\textbf{Capability}  \\ \hline \hline
        \textit{A}    & 3.40GHz 8-Core Intel i7-6700 & 4GB    & Weak           \\ 
        \textit{B}    & 3.40GHz 8-Core Intel i7-6700 & 8GB    & Moderate       \\ 
        \textit{C}    & 3.70GHz 16-Core Xeon W-2145 & 32GB   & Powerful       \\
        \hline
    \end{tabular}
\end{table}

\begin{table}[t] 
  \caption{Statistics of used graph datasets.}
  \label{table:dataset}
  \centering
  \setlength{\belowcaptionskip}{-0.5cm}
  \begin{tabular}{|c|c|c|c|c|c|}
    \hline 
    \textbf{Dataset} & \textbf{Vertex} & \textbf{Edge} & \textbf{Feature} & \textbf{Label} & \textbf{Duration} \\ \hline \hline
    SIoT & 16216 & 146117 & 52 & 2 & 1\\ 
    Yelp & 10000 & 15683 & 100 & 2 & 1\\ 
    PeMS & 307 & 340 & 3 & N/A & 12\\ \hline
    \hline
    RMAT-20K & 20.0K & 199K & 32 & 8 & 1\\
    RMAT-40K & 40.0K & 799K & 32 & 8 & 1\\
    RMAT-60K & 60.0K & 1.79M & 32 & 8 & 1\\
    RMAT-80K & 80.0K & 3.19M & 32 & 8 & 1\\
    RMAT-100K & 100K & 4.99M & 32 & 8 & 1\\\hline
    \end{tabular}
\end{table}

\subsection{Performance Comparison}
\label{sec:perf_comparison}
This subsection compares Fograph with the state-of-the-art cloud serving and the straw-man fog approach in metrics of latency, throughput, and inference accuracy, using a cluster of six fog nodes including types of 1$\times$\textit{A}, 4$\times$\textit{B} and 1$\times$\textit{C}.
All types of nodes here run with CPU only since the energy-intensive accelerators are not prevailing on existing IoT computing platforms \cite{xu2021cloud}, though we will still evaluate how GPU improves Fograph’s performance
latter in \S\ref{sec:gpu_enhancement}.

\begin{figure}[t]
  \centering
  \includegraphics[width=0.95\linewidth]{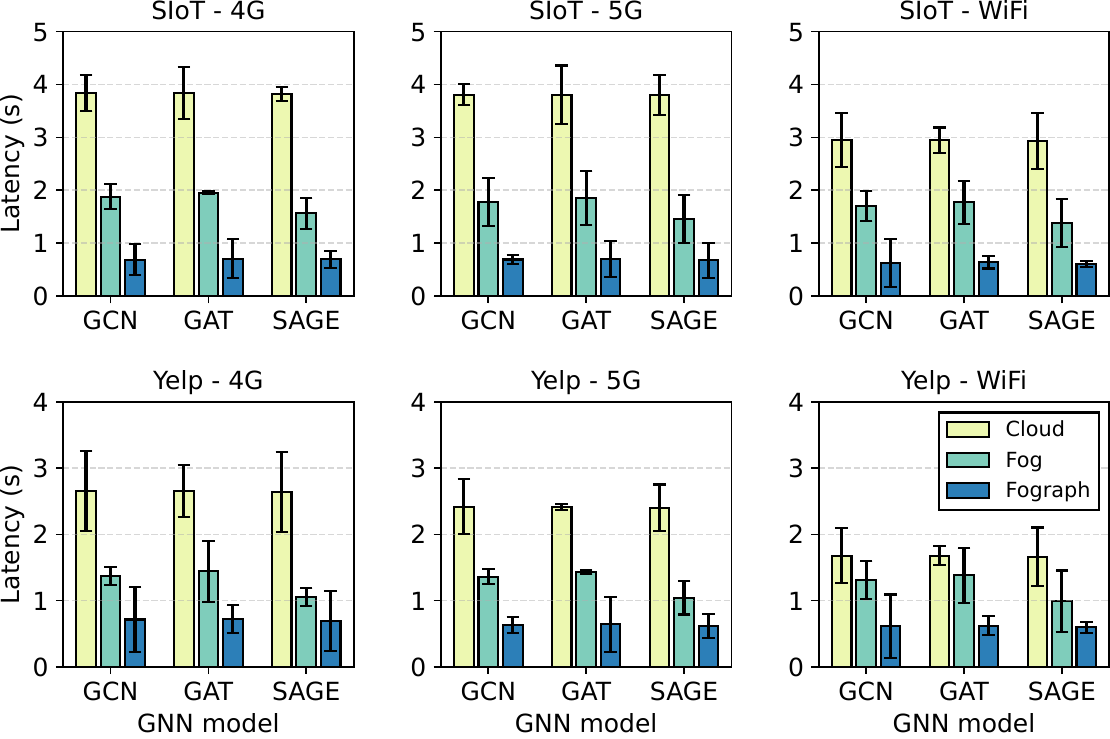}
  \caption{Achieved latency of three GNN models for SIoT and Yelp under varying network conditions.}
  \label{fig:latency_compare}
\end{figure}

\begin{figure}[t]
  \centering
  \includegraphics[width=0.95\linewidth]{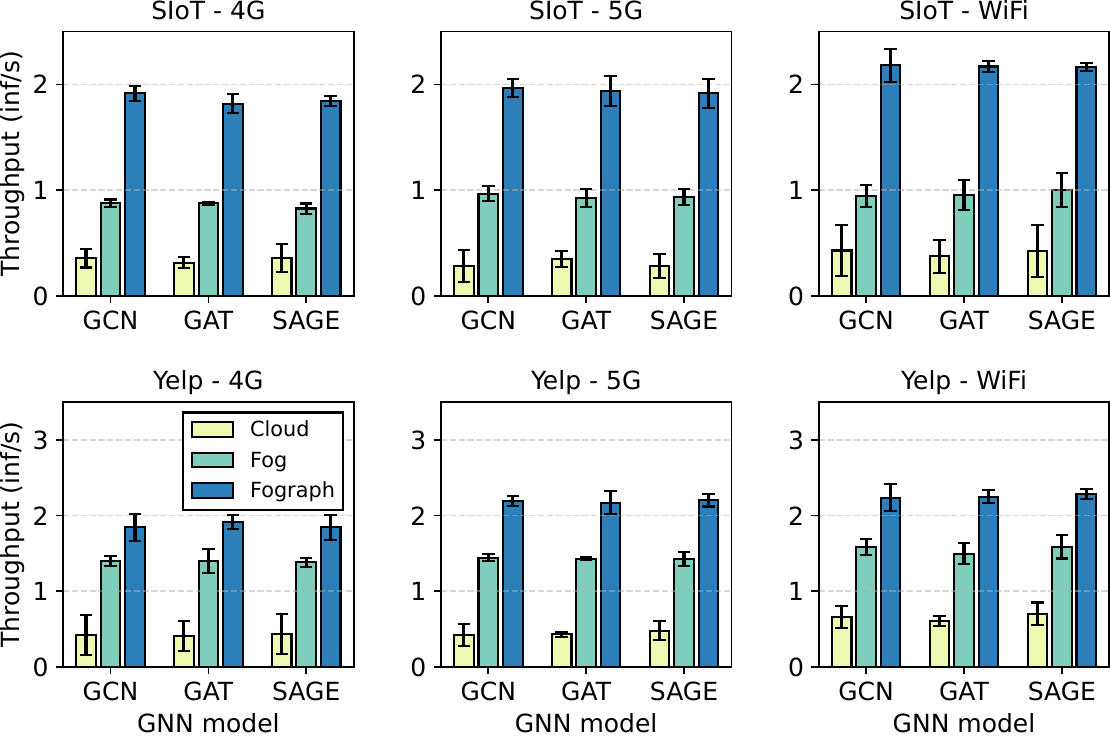}
  \caption{Achieved throughput of three GNN models  for SIoT and Yelp under varying network conditions.}
  \label{fig:throughput_compare}
\end{figure}

\textbf{Latency and throughput.}
Fig. \ref{fig:latency_compare} and Fig. \ref{fig:throughput_compare} respectively show the achieved inference latency and throughput across varying models, datasets and network conditions.
Given a dedicated dataset and network speed, \textit{e.g.} the upper left subfigure SIoT-4G, the cloud serving yields the highest latency and the lowest throughput despite the models because it is constrained by the inevitable communication overhead of remote transmission.
The fog approach significantly lessens the latency compared with cloud, showing that trivially applying fog computing still enjoys its architectural advantages.
More specifically, by shortening the transmission distances between data sources and processing nodes, the fog-based solutions can significantly alleviate the communication bottleneck of centralized cloud serving caused by the long-tail data collection, since GNN inferences can be executed only if all graph data arrive.
Fograph further shrinks the costs, achieving $<$1s inference latency, realizing efficient serving in practical sense.

Across all setups, Fograph consistently delivers the highest performance with a latency reduction up to 82.18\% and 63.70\%, and a throughput improvement of 6.84$\times$ and 2.31$\times$, all over cloud and fog, respectively. 
This can be attributed to our heterogeneity-aware IEP and communication optimizer on maximizing the computing resource utilization while minimizing data uploading overhead.
Traversing horizontally the subfigures with varying network conditions, we observe that poorer channels induce higher speedups of Fograph, \textit{e.g.} averagely from 4.67$\times$ to 5.39 $\times$ on SIoT over cloud when switching WiFi to 4G.
Vertically varying the datasets, a larger source graph enhances Fograph's superiority in saving costs, \textit{e.g.} an average latency reduction of 80.63\% and 70.21\% over cloud for SIoT (larger) and Yelp (smaller), respectively.

An interesting observation of Fig. \ref{fig:latency_compare} is that the serving latency seems to be relatively stable when the dataset and network condition are determined notwithstanding the models.
It is interpreted that the communication overhead majorly dominates the total cost in the cases of GCN, GAT and GraphSAGE inferences.
Although execution's impact will be promoted if larger GNN models are applied, this fact highlights the ponderance of communication optimization in end deployment, which is exactly what Fograph emphasizes.
In Fig. \ref{fig:throughput_compare}, the superiority of Fograph is more evident in contrast with the baselines, validating the effectiveness of optimizing bandwidth utilization and pipeline execution.

\textbf{Accuracy.}
While Fograph profitably exploits the compression technique in reducing data collection overhead, its practicability yet relies on inference accuracy.
To investigate that, we configure the communication optimizer with default settings and assess the approaches under WiFi connections\footnote{The inference accuracy is irrespective to the network conditions. The data corruption caused by unstable transmission, which may decrease the accuracy, is not considered in this paper.}. 
Table \ref{table:accuracy} shows the inference accuracy for SIoT and Yelp upon three models.
The cloud and fog approaches maintain the same accuracy as they both reserve full precision features.
Fograph drops $<$0.1\% for both SIoT and Yelp, which will not cause substantial usability issues on the numerical results.
The accuracy loss is minimal in SIoT because its features are simply one-hot encoded, where the outcome of quantization and compression is maximized without side effects.

\begin{table}[t]
    \centering
    \caption{Inference accuracy on SIoT and Yelp.}
    \label{table:accuracy}
    \begin{tabular}{|c|ccc|ccc|}
        \hline
        \multirow{2}{*}{\textbf{Method}} & \multicolumn{3}{c|}{\textbf{SIoT (\%)}} & \multicolumn{3}{c|}{\textbf{Yelp (\%)}} \\ \cline{2-7} 
                & \textbf{GCN}   & \textbf{GAT}   & \textbf{SAGE}  & \textbf{GCN}   & \textbf{GAT}   & \textbf{SAGE}  \\ \hline \hline
        Cloud   & 89.98 & 86.08 & 95.50 & 92.19 & 86.30 & 91.73 \\
        Fog     & 89.98 & 86.08 & 95.50 & 92.19 & 86.30 & 91.73 \\
        Fograph & 89.97 & 86.08 & 95.48 & 92.12 & 86.20 & 91.70 \\ \hline
    \end{tabular}
\end{table}

\begin{figure*}
	\subfigure[Data placement.]{
        \begin{minipage}[t]{0.21\linewidth}
            \centering
            \includegraphics[height=2.8cm]{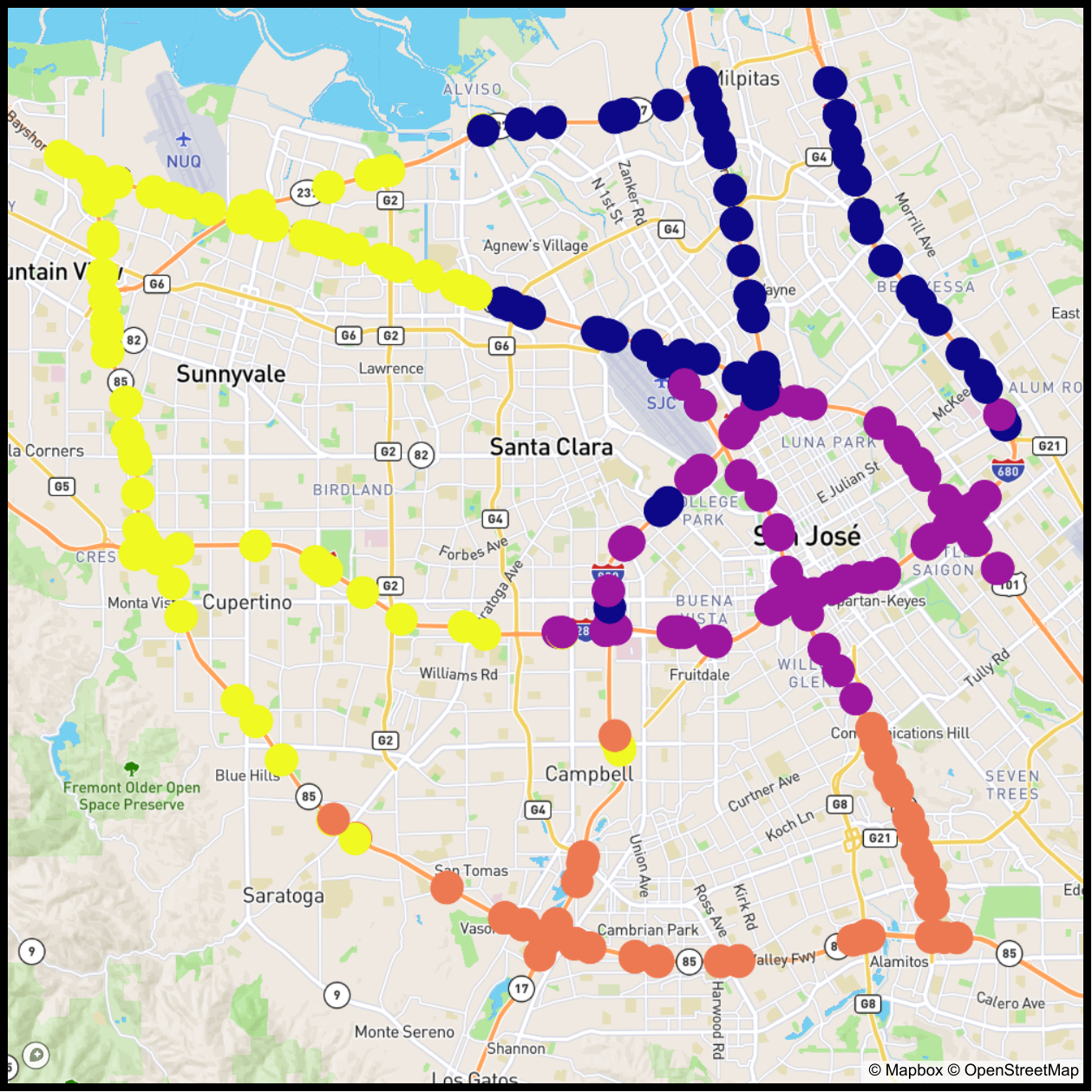}
            \label{fig:case_map}
        \end{minipage}
	}
	\quad
	\subfigure[Load distribution.]{
        \begin{minipage}[t]{0.21\linewidth}
            \centering
            \includegraphics[height=2.8cm]{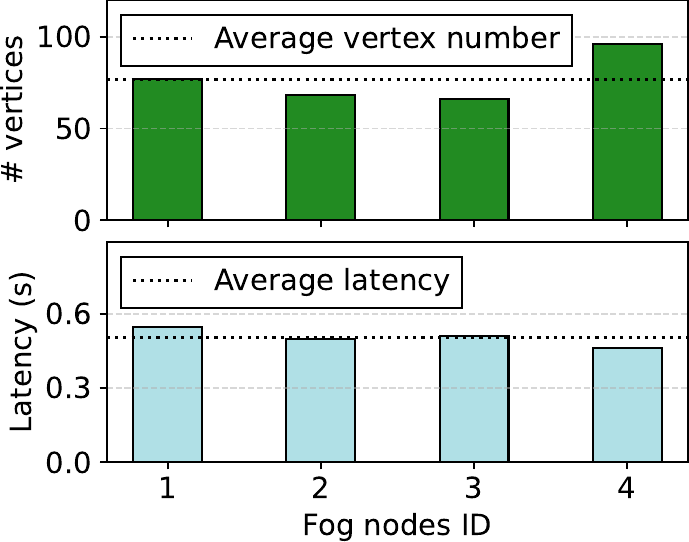}
            \label{fig:case_load}
        \end{minipage}
	}
    \quad
    \subfigure[Latency comparison.]{
        \begin{minipage}[t]{0.21\linewidth}
            \centering
            \includegraphics[height=2.8cm]{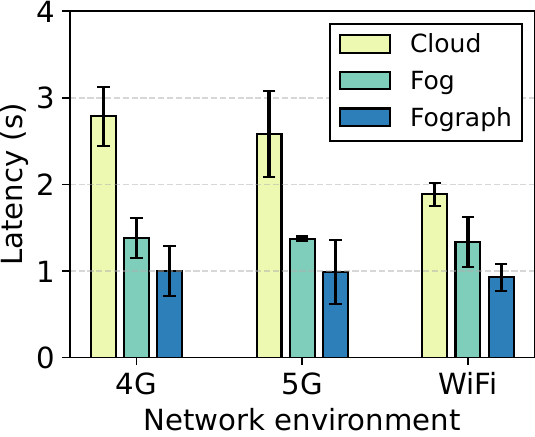}
            \label{fig:case_latency}
        \end{minipage}
	}
    \quad
    \subfigure[Throughput comparison.]{
        \begin{minipage}[t]{0.21\linewidth}
            \centering
            \includegraphics[height=2.8cm]{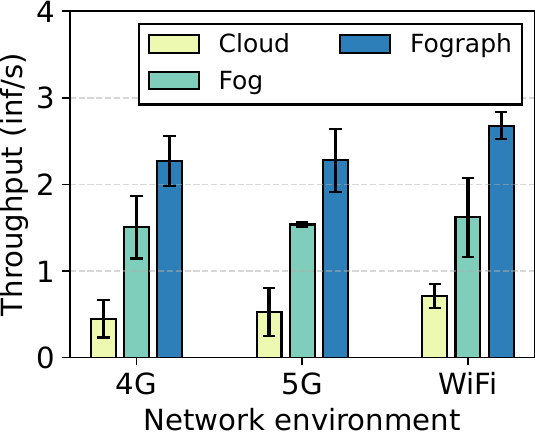}
            \label{fig:case_throughput}
        \end{minipage}
	}
	\caption{Case study results, where (a) visualizes the PeMS sensors' spatial distribution and placement on fog nodes; (b) shows the number of assigned vertices and inference latency measurements of fog nodes; (c) and (d) presents the latency and throughput results, respectively.}
	\label{fig:case_study}
\end{figure*}

\subsection{Case Study: Traffic Flow Forecasting}
\label{sec:case_study}

This subsection uses a realistic case to complement the performance comparison in demonstrating Fograph's superiority.
We emulate a traffic flow forecasting application \cite{guo2019attention} by running the inferences to predict future flow over the PeMS sensor network in an expected time window.
In this setting, four nodes are employed: 1$\times$\textit{A}, 2$\times$\textit{B}, and 1$\times$\textit{C}.
The reason behind configuring such a smaller, weaker cluster (minus two Type-\textit{B} fogs from the testbed in \S\ref{sec:perf_comparison}) is to match the computing resources with the much smaller graph dataset targeted in this experiment (compared to SIoT and Yelp).
To fit the application, we select ASTGCN, a representative spatial-temporal model with GCN as building blocks.

\textbf{IEP result.}
Fig. \ref{fig:case_map} visualizes the sensors' spatial distribution of PeMS and its data placement results.
Each vertex is colored to indicate its placed fog node.
From a global view, the sensors, as the graph vertices, exhibit a clustering pattern of their placement, demonstrating the locality preservation of IEP.
Moreover, the different vertices number of each partition implies its heterogeneity awareness.
This can be clearer in Fig. \ref{fig:case_load}, which counts the number of assigned vertices and the execution latency of each fog node.
The execution latency results among fog nodes are well close though they hold entirely different numbers of vertices.
In particular, the 4-th fog node (Type \textit{C}) accommodates the most vertices (the yellow points in Fig. \ref{fig:case_map}), but conversely takes the lowest execution time.
This is because that it possesses the highest computing capability among others (Type \textit{A} and \textit{B} fog nodes), and thus Fograph enforces it to afford more workload (vertices).
Such load balance attributes to the heterogeneity-aware IEP, which effectively aligns execution workload to the diverse computing resources and ensures maximum resource utilization towards parallel and distributed inference.

\textbf{Latency and throughput.}
Fig. \ref{fig:case_latency} shows the inference latency and throughput of the approaches under 4G, 5G, and WiFi.
Analogous to the comparison in Fig. \ref{fig:latency_compare}, Fograph surpasses the baselines with the lowest costs.
Specifically, with varying channels, Fograph consistently attains the lowest serving latency, achieving speedups of up to 2.79$\times$ and 1.43$\times$ over cloud and fog, respectively.
The latency of traffic flow forecasting appears higher cost than the SIoT and Yelp tasks because it predicts a window of 12 flow possibilities for every 5 minutes in the next hour, which puts a higher workload on execution.
Fig. \ref{fig:case_throughput} presents the corresponding throughput results, where Fograph outperforms all other approaches, showing its superiority.

\textbf{Forecasting performance.}
Table \ref{table:traffic_acc} records three common evaluated metrics for traffic flow forecasting: Mean Absolute Error (MAE), Root Mean Square Error (RMSE) and Mean Absolute Percentage Error (MAPE).
We compare Fograph in two time horizons against cloud, straw-man fog, and an additional compression method that uniformly compresses all feature vectors into 8-bit precision.
Similar to the results in Table \ref{table:accuracy}, Fograph induces minimal error expansion of around 0.1 for all metrics compared with the full precision version (cloud and fog).
In contrast, uniformly quantizing all features to 8-bit results in an evident error gap to Fograph, which may significantly sacrifice the serviceability of the forecasting results.
Such prediction advantages of Fograph origins from our degree-aware quantization choice on exploiting the data sensitivity in GNN inference, enabling both latency reduction and accuracy reservation.

\begin{table}[t] 
    \centering
    \caption{Traffic flow forecasting errors.}
    \label{table:traffic_acc}
    \begin{tabular}{|c|ccl|ccl|}
        \hline
        \multirow{2}{*}{\textbf{Method}} & \multicolumn{3}{c|}{\textbf{15min}} & \multicolumn{3}{c|}{\textbf{30min}} \\ \cline{2-7} 
                                & \textbf{MAE}     & \textbf{RMSE}    & \textbf{MAPE}    & \textbf{MAE}     & \textbf{RMSE}    & \textbf{MAPE}    \\ \hline \hline
        Cloud                   & 17.71   & 29.92   & 11.84   & 18.66   & 30.97   & 12.27   \\
        Fog                     & 17.71   & 29.92   & 11.84   & 18.66   & 30.97   & 12.27   \\
        Fograph                 & 17.75   & 30.05   & 11.93   & 18.73   & 31.12   & 12.38   \\
        Uni. 8-bit           & 18.79   & 30.26   & 12.97   & 19.74   & 32.01   & 13.38   \\ \hline
    \end{tabular}
\end{table}

\subsection{Optimization Implication}
\label{sec:implication}

This subsection investigates the performance boost of each individual optimization technique introduced in \S\ref{sec:design}, using SIoT dataset and the cluster described in \S\ref{sec:case_study}.

\begin{figure}[t]
  \centering
  \includegraphics[width=0.9\linewidth]{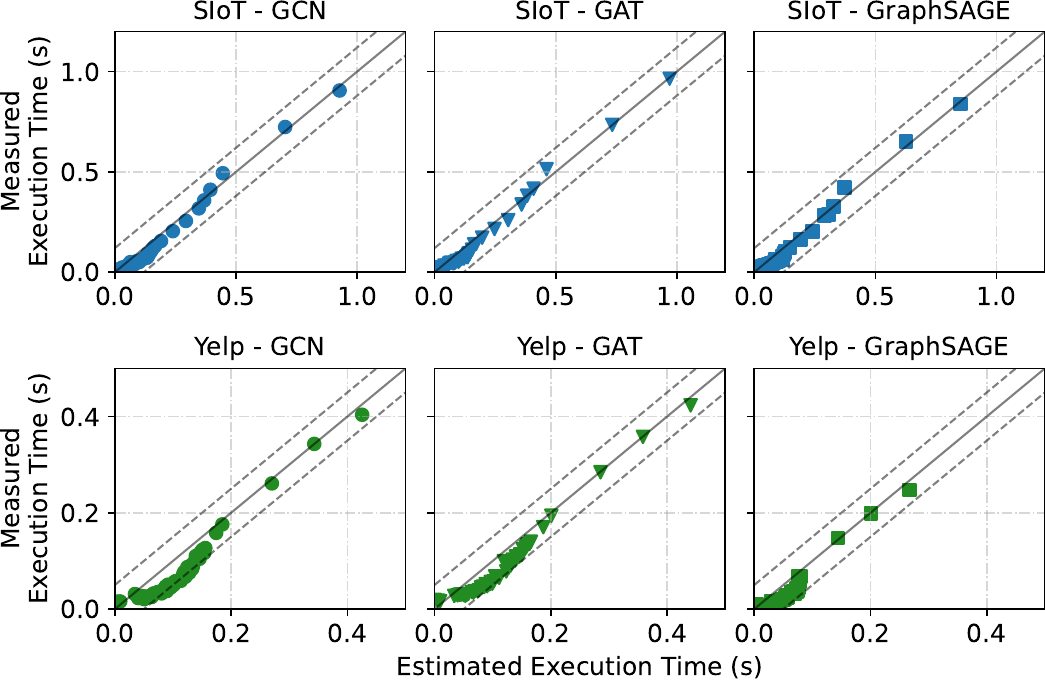}
  \caption{Profiling measurements and the prediction results of varying GNN models and datasets. The solid line indicates the estimated execution time exactly equals the actual execution time, whereas the dashed lines draw a relative difference of $\pm$10\% between them.}
  \label{fig:profiling}
\end{figure}

\textbf{Metadata profiler.}
We first show the profiler's effectiveness by comparing the estimated execution time predicted by the profiler with the real execution time measured in actual inferences.
Fig. \ref{fig:profiling} depicts the profiling and prediction results for different GNN models and different datasets, collected on a fog node of type $B$.
The solid line indicates an exact equivalence between the measurement and the prediction, while the dashed lines mean a relative difference of $\pm$10\% between them.
All data points are encompassed by the dashed lines, which implies a small and bounded variance between actual execution latency and estimates, demonstrating the effectiveness of the proxy-guided profiling methodology.
Moreover, the figure shows that if a graph has a higher execution latency than the other one, their latency estimates still preserve the ordering.
This demonstrates that the profiler's latency prediction is an appropriate tool to guide IEP's data placement optimization.

\begin{figure}[t]
  \centering
  \includegraphics[width=0.9\linewidth]{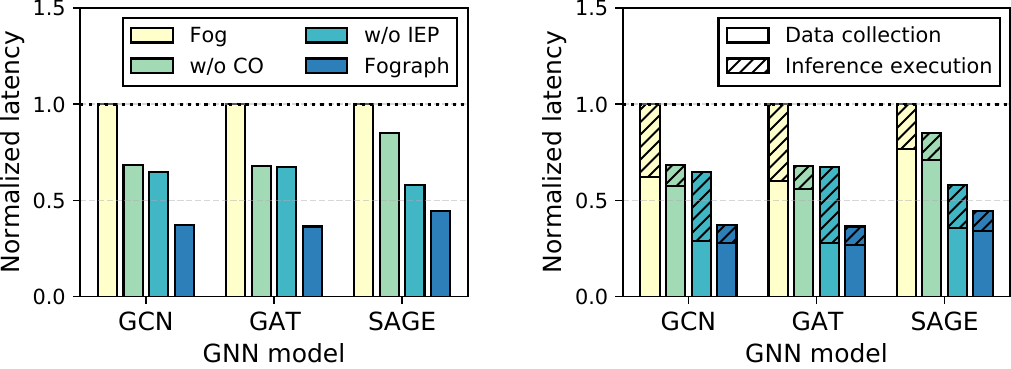}
  \caption{Performance comparison of ablated Fograph counterparts in terms of inference execution planner (IEP) and communication optimizer (CO). }
  \label{fig:implication}
\end{figure}

\textbf{Inference execution planner and communication optimizer.}
Fig. \ref{fig:implication} normalizes the latency of four approaches: fog, Fograph, and its ablated variants without inference execution planner (IEP) or communication optimizer (CO).
Specifically, Fograph without IEP replaces its data placement strategies with the one in the straw-man fog approach, and Fograph without CO simply applies direct data transmission between data sources and fog nodes.
All other configurations keep unaltered for these two ablated counterparts except for their respectively targeted modules.
From a global view, we observe that both modules make sense for performance improvement, while a combination promotes higher speedups.
The IEP and CO take similar effects but their focuses are different as indicated by their execution statistics in Fig. \ref{fig:implication} (right).
The former essentially solves a graph layout to maximize the parallelization and thus saves the overall latency on the execution side given a smaller execution ratio, whereas the latter centers around the data uploading aspect and validly reduce the transmission costs, which lowers the proportion of the communication side.
The almost orthogonal optimization directions make their incorporation fully benefit the performance gains.

\begin{figure}[t]
  \centering
  \includegraphics[width=0.88\linewidth]{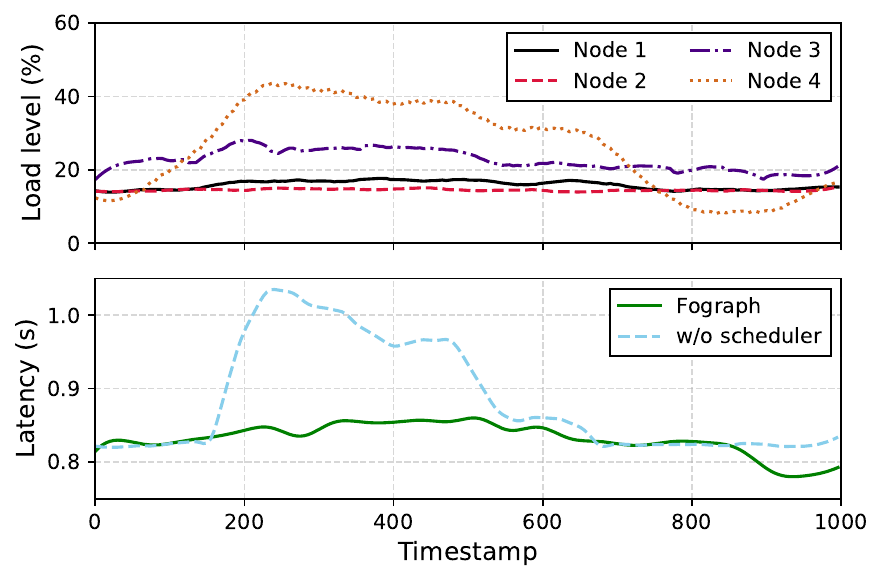}
  \caption{Fograph's behaviour on real load trace, where the top subfigure depicts the load fluctuation of four fog nodes and the bottom subfigure shows the inference latency variation of Fograph and its counterpart without adaptive workload scheduler.}
  \label{fig:cpu_variation}
\end{figure}

\textbf{Workload scheduler.}
To appraise the responsiveness of Fograph in adapting to dynamic load fluctuation, we target production workload traces from Alibaba \cite{weng2022mlaas}. 
The trace contains background CPU load variation running on clusters and we select a snapshot of 1000 timestamps to exert pressures on fog nodes, as shown in Fig. \ref{fig:cpu_variation} (top).
The associated behaviors of Fograph with/without workload scheduler are recorded in Fig. \ref{fig:cpu_variation} (bottom).
Note that the scheduler-less version is a non-dynamic counterpart implemented by deactivating the workload scheduler module in Fograph and maintaining the original IEP data placement all the time.
Given the superiority of IEP, it represents a fair performance produced by non-dynamic load-balancing methods like fixed division of computing pressure according to computing capability and communication distance.
When the fog nodes share similar load levels (timestamp [0, 150]), the two versions report close inference latency.
As node 4's background load increases, the latency of the w/o scheduler version rapidly climbs while the integrated Fograph keeps relatively steady.
Particularly, the serving latency can exceed 1s in the extreme situation without scheduler, while Fograph keeps immersing below 0.9s.
At the other end, when node 4's load diminishes, its computing capability is released so that Fograph is able to further lessen the costs and achieve up to 18.79\% latency reduction over the ablated copy.
Overall, we observe that the lack of scheduler leads to a latency trajectory going after the overloaded nodes' trace changes and magnifies load fluctuation in costs.
Instead, by employing the workload scheduler, Fograph can adjust to resource dynamics and provides continuously stable serving.
Moreover, we observe that Fograph acts agilely to load dynamics for low-latency service provisioning.
With a very mild communication delay of $\sim$0.2s between the metadata server and fog nodes, our measurements report an average of 4.32s from imbalance detection to load migration.

\subsection{Scalability} \label{sec:scalability}

\begin{figure}[t]
  \centering
  \includegraphics[width=0.95\linewidth]{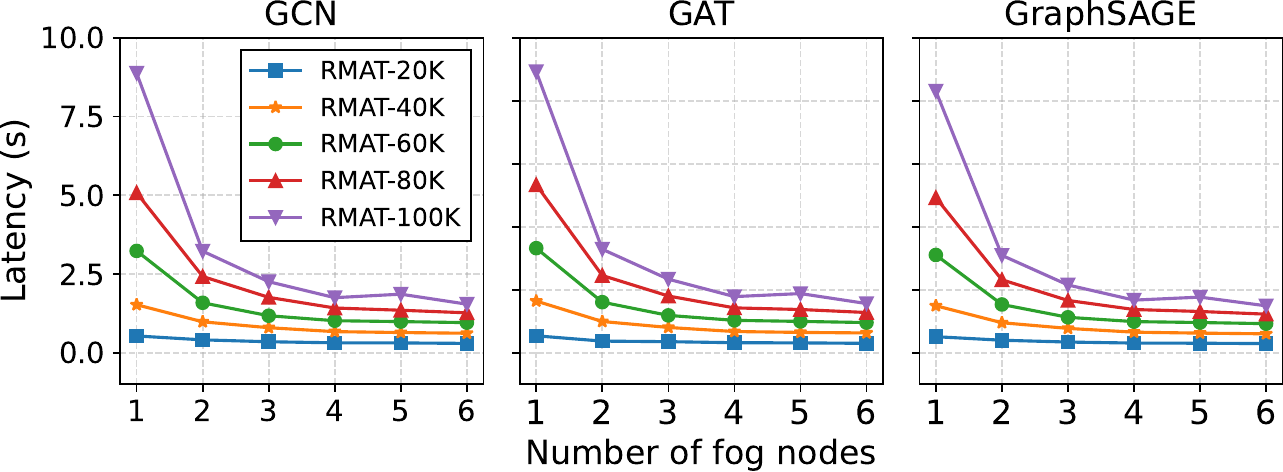}
  \caption{The serving latency of Fograph with varying size of datasets and varying number of fog nodes.}
  \label{fig:scalability}
\end{figure}

This subsection investigates the scalability of Fograph using much larger synthetic RMAT datasets, with varying fog nodes of type $B$.
Fig. \ref{fig:scalability} shows Fograph's serving costs over the models and datasets as the number of fogs increases, where we observe the latency effectively shrinks with the augment of computing resources.
Employing multiple fogs ($>$2) performs much better than merely using one fog for every size of graphs, demonstrating Fograph's efficient resource utilization with parallel and distributed execution.
Upon larger graphs (\textit{e.g.} RMAT-100K), Fograph can gain clearly much performance improvement when appending additional fog nodes, suggesting Fograph's capability in handling heavier serving workload.
We remark that the serving costs converge for all graphs when there are ample fog nodes, which implies that Fograph can readily afford million-edge graphs with just six moderate fog nodes.

\subsection{GPU Enhancement} \label{sec:gpu_enhancement}
\begin{figure}[t]
  \centering
  \includegraphics[width=0.78\linewidth]{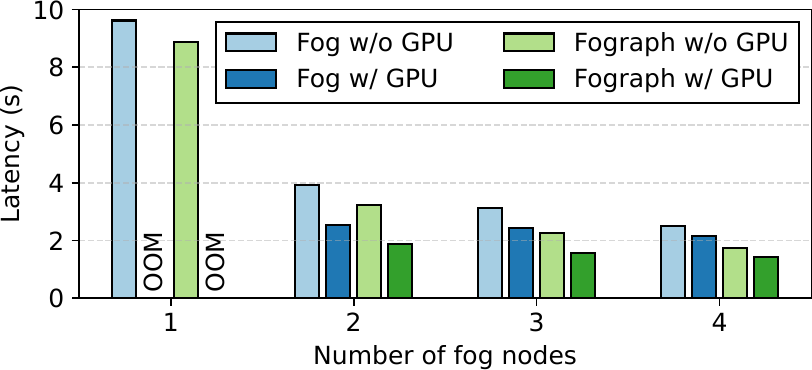}
  \caption{The serving latency of varying number of fog nodes with and without GPU. OOM indicates the out-of-memory error.}
  \label{fig:gpu_latency}
\end{figure}

This subsection explores how GPU enhances Fograph towards serving performance.
We equip each fog node of type \textit{B} with an Nvidia GTX 1050 GPU, and run GCN inference over the synthetic RMAT-100K dataset.
Fig. \ref{fig:gpu_latency} visualizes the achieved latency of the straw-man fog method and Fograph, with and without GPU.
For a single fog, both fog and Fograph fail with GPU and encounter the out-of-memory (OOM) error, due to the limited GPU memory.
By extending to multi-fog, however, GPU clearly shows its advantage over CPU, and makes progressive improvements as the number of fogs grows, demonstrating Fograph's enhanced performance with hardware accelerators.
We also find that serving with a small number of fog nodes (\textit{e.g.} with 2 fog nodes) exhibits a broader performance gap between solutions with and without GPU.
This implies GPU is particularly expedient when the available fog resource is inadequate towards the targeted GNN services.
Moreover, Fograph without GPU can reach even lower latency compared with the straw-man fog with GPU (with $>$3 fog nodes), proving the remarkable performance advance from the proposed set of optimizations.

\section{Related Work}
\label{sec:related_work}

As an intersection of GNN processing and fog infrastructure, Fograph provides a group of GNN-generic designs for maximizing the architectural advantages of fog computing.
Next we discuss these two lines.

\textbf{Accelerating GNN processing.}
A growing body of recent worksfrom both the research \cite{jia2020improving, tian2020pcgcn, liu2020g3, thorpe2021dorylus, mohoney2021marius, cai2021dgcl} and industry communities  \cite{ma2019neugraph, zhu2019aligraph, zhang2020agl, gandhi2021p3, wang2021flexgraph, zheng2022bytegnn}  have focused on reaching high performance GNN processing by optimizing different execution levels.
From a hardware perspective, a number of domain-specific processors \cite{hu2020featgraph, ma2019neugraph, geng2020awb, kiningham2020greta} are designed by either customizing the microarchitectures to the GNN execution semantics or alternating hardware-software interfaces for graph data reading/writing.
The aggregate and update functions are discriminatingly optimized according to their access patterns, data reusability, and computation intensities.
Fograph is orthogonal to these fundamental efforts and can directly enjoy their pedestal improvements.

From a library perspective, PyG \cite{fey2019fast} and DGL \cite{wang2019deep} are representative efforts that provide GNN-tailored API supports atop the neural networks execution engines like PyTorch and TensorFlow.
The message-passing model is exploited for unified programmability and the matrix operations are particularly optimized with specialized kernels.
Fograph utilizes these libraries as the backend so as to fully benefit from their underlying optimizations.

From a system perspective, GNN execution is treated and optimized as an integrated production process considering its global lifecycle from graph storage \cite{zhu2019aligraph, zhang2020agl}, data loading \cite{lin2020pagraph, min2021large}, memory management \cite{jia2020improving, jia2020redundancy, wu2021seastar} to multi-GPU execution \cite{ma2019neugraph, kim2021accelerating}.
Miscellaneous systems are proposed to address some of the aspects, \textit{e.g.}, $P^3$ \cite{gandhi2021p3} for scalable processing and Dorylus \cite{thorpe2021dorylus} for affordable training.
Nonetheless, a majority of the systems focus only on training rather than inference, and all are on single powerful machines or cloud environments, ignoring the inherent data uploading overhead in the serving pipeline.
Instead, Fograph capitalizes the more realistic and practical scenarios of GNN serving, emphasizes and implements the unique potential of fog computing to yield high performance.

\textbf{Fog-enabled DNN inference.}
To achieve efficient intelligent services close to the end, a line of works \cite{mao2017modnn, zeng2019boomerang, zhao2018deepthings, li2019edge, hadidi2018distributed, zeng2020coedge} have explored collaborative execution with the assist of fog computing.
By splitting and mapping the input workloads or model parameters, these works parallelize CNN or RNN inferences over a cluster of fog nodes to meet dedicated service-level objectives.
A few constraints are usually added in accordance with applications requirements such as execution deadline \cite{zeng2020coedge, laskaridis2020spinn} and energy efficiency \cite{hadidi2018distributed}.
Fograph also lies in this category of work by first bringing GNN workload into the fog-enabled intelligence.
Further, Fograph tailors its design to bridge the gap between the unique characteristics of GNN execution and the distributed and heterogeneous environments of fog, which notably improves the overall performance beyond cloud and vanilla fog deployment.

\section{Discussion and Future Work}
\label{sec:discussion}

Fograph is a pilot effort in bringing fog computing to GNN processing systems and yet has certain limitations.

\textbf{Exploiting inference context.}
Fograph has explored leveraging the unique characteristics of GNN and graph, \textit{e.g.} degree-aware quantization, to boost the serving performance.
As a result, Fograph's performance largely relies on the size and complexity of input data volume, where input graph data with larger vector sizes and sparser features can further highlight its superiority over cloud solutions.
However, the rich semantics of other inference contexts still call for exploitation including input graph properties, application workflow patterns, and inference functions specialties.
Recent works on GNN performance characterizations \cite{wang2020gnnadvisor, huang2021understanding} can provide useful insights to enhance Fograph and guide the parameters tuning and module altering.

\textbf{Complex heterogeneity and dynamics.}
The heterogeneity and dynamics considered in Fograph are mainly on the execution side, \textit{i.e.} the diverse and fluctuated computing capabilities.
The allocated bandwidth is implicitly assumed to be relatively stable for all fog nodes. 
Although it is realistic in many fog computing cases (\textit{e.g.} in a closed factory or campus \cite{porambage2018survey}), real-world serving could confront more complicated situations where the connections between sensors and fog nodes are varied and even fail \cite{almeida2021dyno}.
In these cases, we may ignore those few vertices with ultra-high transmission delay during data collection in order to stabilize the overall serving latency.
Consequently, the long-tail distribution of data collection time is squeezed and cloud serving could act as an efficient complement to Fograph for more robust serving performance.

\textbf{Scalability and fog-cloud collaboration.}
Fograph's scheduling employs a centralized metadata server to attain fair resource utilization, which in principle tunes the communication-computation tradeoff for the distributed runtime.
While it works well for typical moderate-scale deployment in fog scenarios, it may fall short in scaling up with very huge graphs and massive fog nodes due to the single-point scheduler.
To address that, one of the potential solutions is to apply a lightweight, decentralized data placement strategy for inference execution planning such as hashing \cite{lin2020pagraph,gandhi2021p3}.
Alternatively, we may utilize the abundant cloud resources as a supplement to accomplish scalable GNN processing, \textit{e.g.} by designing a fog-cloud collaboration mechanism that uses fog nodes to collect and compress data and cloud servers to accommodate full-batch GNN processing at scale.

\textbf{Other optimizing objectives. }
The proposed system concentrates on rendering GNN model serving in a real-time manner, whereas end deployment may tackle additional Service-Level Agreements (SLAs) like server costs and memory footprint \cite{laskaridis2020spinn, zeng2022gnn}.
The deadline can be integrated as a constraint in the workload scheduler, while the memory issue can be accounted for by redesigning the problem formulation in IEP. 
Composite SLAs require supplementary scheduling in jointly optimizing the system behaviors with multiple objectives.
Our future work intends to enhance Fograph to meet additional types of objectives, \textit{e.g.} energy consumption and privacy preservation.

\section{Conclusion}
\label{sec:concolusion}

In this paper, we present Fograph, a distributed GNN inference system that addresses real-time GNN serving with fog computing.
Fograph introduces a brand new serving pipeline that allows exploiting the diverse resources from distributed fog servers for resilient service rendering.
Through a heterogeneity-aware inference execution planner and adaptive workload scheduler that effectively maps the input graph over multiple fog nodes, Fograph can maximize the parallelization while simultaneously adapting to resource fluctuation.
By employing a GNN-specific communication optimizer, Fograph is able to deliver higher performance over the state-of-the-art cloud serving and basic fog deployment, without sacrificing the overall system’s accuracy and validity.
Since GNNs have been widely adopted in a broad range of IoT and fog scenarios, the proposed system and its workflow can serve as a basis for future analysis and optimization on specific GNN-based services.

\ifCLASSOPTIONcaptionsoff
  \newpage
\fi

\bibliographystyle{IEEEtran}
\bibliography{main.bib}

\appendix

\section{Mathematical Proofs}

\subsection{Proof of Theorem 1}
\label{sec:proof_1}
\begin{proof}
We prove Theorem 1 by reducing $\mathcal{P}$ to the \textit{Minimum Makespan Scheduling Problem} (\textit{MMSP}).
First, we consider a special case of $\mathcal{P}$ by forcing the fogs' allocated bandwidth and computing capability to be identical.
Suppose an input graph that comprises isolated vertices, to collect and compute inference of these vertices can thus be regarded as individual jobs.
Regarding $\mathcal{P}$, the objective is to assign the vertices to the fogs such that the maximum job processing time on fogs is minimized, which is exactly a \textit{MMSP} problem.
Given that \textit{MMSP} has been proved to be NP-hard when there are two or more machines \cite{bruno1974scheduling}, $\mathcal{P}$ is NP-hard when $n \geq 2$, where $n$ is the number of fogs.
\end{proof}

\subsection{Proof of Theorem 2}
\label{sec:proof_2}
\begin{proof}
We give the compression ratio of \textit{degree-aware quantization} (DAQ) by treating a vertex's degree as a discrete random variable $D$.
First, we build the degree distribution of a given graph $\mathcal{G} = (\mathcal{V}, \mathcal{E})$ by counting the frequency of its vertices' degrees, and subsequently obtain a \textit{Cumulative Distribution Function} (CDF) $F_D(d)$ with respect to vertex degree.
Formally, the CDF defines the probability that the vertex degree $D$ takes a value less than or equal to a given threshold $d$: 
\begin{align}
    F_D(d) = \mathbf{P}(D \leq d). 
\end{align}

Using CDF, we can derive the percentage $r_i$ of vertices that locate in the four intervals divided by the DAQ triplet $\langle D_1, D_2, D_3 \rangle$:
\begin{align}
    r_{i} = F_D(D_{i+1}) - F_D(D_{i}), \ i \in \{0, 1, 2, 3\}. \label{eq:number_vertex}
\end{align}
where we supplement that $D_0 = 0$ and $D_4 = D_{\text{max}}$. 
Particularly, remark that $F_D(0) = 0$ and $F_D(D_{\text{max}}) = 1$, thus we have
\begin{align}
    r_0 &= F_D(D_1) - F_D(D_0) = F_D(D_1), \label{eq:N_0} \\
    r_3 &= F_D(D_4) - F_D(D_3) = 1 - F_D(D_3). \label{eq:N_3}
\end{align}

The total bits $B^{\text{quant}}$ of the quantized feature vectors is $\sum_i r_i |\mathcal{V}|q_i$, and substitute Equation (\ref{eq:number_vertex}) we have
\begin{align}
    B^{\text{quant}} = \sum_i [F_D(D_{i+1}) - F_D(D_{i})]|\mathcal{V}|q_i, \ i \in \{0, 1, 2, 3\}.
\end{align}

With Equations (\ref{eq:N_0}) and (\ref{eq:N_3}), we extrapolate the above $B^{\text{quant}}$ to
\begin{align}
    B^{\text{quant}} = |\mathcal{V}|[q_3 - \sum_i F_D(D_i)(q_{i} - q_{i-1})], \ i\in\{1,2,3\}.
\end{align}

Given the original feature bitwidth $Q$, the overall bitwidth $B^{\text{origin}}$ of $\mathcal{G}$'s original features is $|\mathcal{V}|Q$, and hence the compression ratio directly follows
\begin{align}
    \frac{B^{\text{quant}}}{B^{\text{origin}}} = \frac{q_3}{Q} - \frac{1}{Q}\sum_i F_D(D_i)(q_i - q_{i-1}), \ i\in\{1,2,3\}.
\end{align}
\end{proof}

\section{Details of The Used Datasets}
\label{sec:dataset}

\subsection{Details of Real-World Datasets}
\label{sec:real_world_datasets}
Three real-world graph datasets are employed in our evaluation.
The first is SIoT \cite{marche2020exploit}, a graph of socially connected Internet-of-Things collected in Santander, Spain.
It includes 16216 devices as vertices with 146117 social links, and each vertex attaches a 52-dimension feature that identifies its properties such as the device's type, brand, and mobility.
Each device is managed by an organization or a person, and thus yields a label of whether public or private.
The GNN serving task over SIoT is to identify the devices' labels, by inferring from their features and relationships.

The second is Yelp, a subgraph extracted from its complete back-up \cite{rayana2015collective}, which collects reviews for a set of hotels and restaurants in Chicago.
A vertex in the Yelp graph is a review comment, represented by a Word2Vec \cite{mikolov2013efficient} vector, and each connection indicates the two reviews share a common history like they are posted by the same user.
The comments are separated into two classes: spam reviews that produce fake and filtered content, and benign reviews that are not filtered.
We run inference on Yelp to identify the spammers.

The third is \textit{PeMS}, which is collected by Caltrans Performance Measurement System \cite{chen2001freeway} in San Francisco Bay Area, containing the traffic sensors' every-30s records on total flow, average speed, and average occupancy.
Its topology is exactly the road network, comprising 307 loop sensors and 340 edges.
Unlike the above two datasets for prediction on a single moment, PeMS associates the task of forecasting every-5min flows in an hour (12 timestamps) and is used in our case study (\S IV-C).

\subsection{Details of Synthetic Datasets}
\label{sec:synthetic_datasets}
The synthetic datasets are created by RMAT \cite{chakrabarti2004r}, a widely-adopted graph generator that is able to quickly generate realistic graphs.
Specifically, we set the number of vertices in \{20K, 40K, 60K, 80K, 100K\}, respectively.
To capture the sparsity in realistic graphs, we use the graph density of SIoT, 0.11\%, to steer the generation of edges, and accordingly produce different edge numbers of \{199K, 799K, 1.79M, 3.19M, 4.99M\}.
In addition, we use Node2Vec \cite{grover2016node2vec} to generate the vertices features in 32 dimensions, and community clustering to induce vertices' labels in 8 classes.

\end{document}